\documentclass[prb,twocolumn,aps,showpacs,10pt]{revtex4}
\usepackage{bm}% bold math
\usepackage{color}
\usepackage{amsmath}
\usepackage{amssymb}
\usepackage{wrapfig}
\usepackage{enumerate}
\usepackage{setspace}
\usepackage[hypertex]{hyperref}
\usepackage[caption=false]{subfig}
\usepackage{bbm}
\usepackage{graphics}
\usepackage{graphicx}% Include Fig. files
\usepackage{stmaryrd}

\newcommand{\bra}[1]{\langle #1 |}
\newcommand{\ket}[1]{| #1 \rangle}
\newcommand{\block}[2]{ \begin{array}{#1} #2 \end{array}}

\newcommand{\str}{\,\textrm{str}\,}
\newcommand{\sdet}{\textrm{sdet}\,}
\newcommand{\diag}{\textrm{diag}\,}

\definecolor{gray}{rgb}{0.7,0.7,0.7}

\begin{document}

\title{Supersymmetry approach to delocalization transitions in a network model of the weak field quantum Hall effect and related models.}

\author{S. Bhardwaj}
\affiliation{The James Franck Institute and the Department of Physics, The University of Chicago, Chicago, Illinois 60637, USA}

\author{V. V. Mkhitaryan}
\affiliation{Ames Laboratory, Iowa State University, Ames, Iowa 50011, USA}

\author{I. A. Gruzberg}
\affiliation{Department of Physics, The Ohio State University, Columbus, OH 43210, USA}

\begin{abstract}

We consider a recently proposed network model of the integer quantum Hall (IQH) effect in a weak magnetic field. Using a supersymmetry approach, we reformulate the network model in terms of a superspin ladder. A subsequent analysis of the superspin ladder and the corresponding supersymmetric nonlinear sigma model allows us to establish the phase diagram of the network model, and the form of the critical line of the weak-field IQH transition. Our results confirm the universality of the IQH transition, which is described by the same sigma model in strong and weak magnetic fields. We apply the suspersymmetry method to several related network models that were introduced in the literature to describe the quantum Hall effect in graphene, the spin-degenerate Landau levels, and localization of electrons in a random magnetic field.

\end{abstract}


\date{June 4, 2014}

\maketitle

\section{Introduction}

Anderson localization of a quantum particle or a classical wave in a random environment [\onlinecite{Anderson58}] is a vibrant research field [\onlinecite{AL50}]. One of its central research directions is the physics of Anderson transitions [\onlinecite{evers08}], including metal-insulator transitions and transitions of quantum Hall type (i.e., between different phases of topological insulators). While such transitions are conventionally observed in electronic conductor and semiconductor structures, there is also a considerable number of other experimental realizations actively studied in recent and current works. These include localization of light [\onlinecite{wiersma97}] and microwaves [\onlinecite{microwaves}], cold atoms [\onlinecite{BEC-localization}], ultrasound [\onlinecite{faez09}], and optically driven atomic systems [\onlinecite{lemarie10}].

Especially intriguing is the problem of the plateau transition in the integer quantum Hall (IQH) effect. The nature of the critical state at and the critical phenomena near the IQH transition are at the focus of intense experimental [\onlinecite{Amsterdam-group, Tsui-group, Amado10, Saeed11, Huang12, Shen12}] and theoretical research [\onlinecite{Zirnbauer99, tsvelik, LeClair, Pruisken-Burmistrov, Obuse08b, Evers08b, Slevin09, Burmistrov10, Amado11, BGL, stabilitymap, OBLGE}]. It is worth mentioning here that experiments aimed at understanding the critical behavior near the IQH transition usually study scaling of transport coefficients with temperature. Such scaling inevitably involves the so-called dynamical critical exponent $z$, and can only be explained if one takes into account some mechanism of dephasing of electronic wave functions, such as electron-electron interactions. The subject of the effects of interactions on the IQH transition is important, and we refer the reader to Refs. [\onlinecite{Burmistrov10}, \onlinecite{dephasing}]. One conclusion of this line of research is that if interaction between electrons is short ranged (for example, due to screening), the interaction is irrelevant in the renormalization group sense, and does not change critical properties that do not involve temperature.

Correspondingly, the vast majority of existing theories of the IQH effect focus on models of noninteracting electrons in a strong magnetic field subject to disorder. Depending on the nature of disorder, one can pursue two complementary approaches. The first, a field-theoretic approach, was developed for short-range (Gaussian white-noise) disorder, where the correlation length of the disorder potential is much shorter than the magnetic length. This leads to a nonlinear sigma model with a topological term [\onlinecite{Levine-Libby-Pruisken, Pruisken-84, Pruisken-87, Weidenmuller}]. Khmelnitskii [\onlinecite{Khmelnitskii-83}] and Pruisken [\onlinecite{Pruisken-85}, \onlinecite{Pruisken-87}] argued that the inclusion of the topological term yields a desirable delocalization in the middle of a Landau band, and predicted a two-parameter flow diagram for the diagonal and Hall conductivities. However, it was not possible to extract critical characteristics of the IQH transition from this theory.

A different approach was developed for smooth disorder with correlation length much longer than the magnetic length, and strong magnetic fields, such that $\omega_c\tau \gg 1$, where $\omega_c$ is the cyclotron frequency and $\tau$ the scattering time. In this limit Landau bands are well resolved, and the semiclassical picture is that of the drift motion of an electron along equipotential lines of the disorder potential [\onlinecite{Kazarinovetal}]. Based on this picture of chiral motion in strong magnetic field, Chalker and Coddington (CC) proposed a random network model of the quantum Hall transition [\onlinecite{CC}]. Remarkably, the CC model is very convenient for numerical simulations and captures both qualitative and quantitative aspects of the high-field IQH transition. Moreover, the CC model and its triangular version [\onlinecite{triangular}] admit a (semi) analytical treatment in terms of a real-space renormalization group [\onlinecite{triangular}, \onlinecite{GalRaikh}].

The idea of incorporating disorder via random phases on the links, on which the CC model was based, appeared to be very fruitful and turned the network-model approach into a powerful tool in numerical studies of disordered systems. By imposing proper symmetry requirements on phases on the links and scattering matrices at the nodes, one can build network-model realizations of disordered electronic systems in all 10 symmetry classes of Altland and Zirnbauer [\onlinecite{altland97, zirnbauer96, heinzner05}]. These realizations often involve networks with multiple channels on the links due to higher symmetries of the scattering matrices.

The disorder average in network models can be performed using either the replica method or the supersymmetry (SUSY) method of Efetov [\onlinecite{Efetov}] adapted to networks [\onlinecite{GruzReadSach}, \onlinecite{GruzLudRead}]. In the replica formalism, network models in an anisotropic limit can be mapped to quantum spin chains. [\onlinecite{DHLee}] Similarly, in the SUSY formalism one obtains superspin  chains [\onlinecite{GruzReadSach}, \onlinecite{GruzLudRead}, \onlinecite{MarstTsai}]. A variant of the SUSY method was used to obtain a continuum limit of the CC model [\onlinecite{Zirnbauer}] and to connect the field-theoretical and network-model approaches. The mapping to spin chains provides a way to study network models analytically and numerically. In particular, the superspin chain relevant for the CC model and the IQH transition has been numerically studied by the density matrix renormalization group method in Ref. [\onlinecite{MarstTsai}]. Also, the language of spin chains makes possible to use analogies with more conventional SU(2) spin chains. Later in this paper, we will use intuition gained from the study of SU(2) spin systems. Many such conventional SU(2) spin chains are known to be integrable and amenable to exact solutions by Bethe ansatz. This leads to a natural question as to whether integrability can shed light on the problem of the IQH and other Anderson transitions. Unfortunately, the direct implementation of the SUSY method for the CC model leads to a superspin chain that is not integrable. Attempts to obtain integrable deformations of the CC model and the corresponding superspin chain have been made [\onlinecite{Zirnbauer}, \onlinecite{IkhlevFendleycardy}]. Most likely, these modifications change the critical properties of the models, so their relevance to the IQH transition is not clear at present.

The limit of weak magnetic fields, $\omega_c\tau \ll 1$, is drastically different. In this limit, the electron motion is not purely chiral and the simple picture of the drift motion fails. As a consequence, the CC model is not applicable. The nontrivial behavior of critical states in vanishing magnetic fields, the so-called phenomenon of levitation, was predicted by Khmelnitskii [\onlinecite{Khmelnitskii}] and Laughlin [\onlinecite{Laughlin}] within the field-theoretic (sigma model) approach. They have argued that, when the magnetic field is decreased towards zero, delocalized states float above the Fermi energy. Subsequently, the levitation scenario became an essential component of the global phase diagram of the quantum Hall effect [\onlinecite{GPD}] and was confirmed by a number of experiments [\onlinecite{Exp1, Exp2, Exp3, Exp4, Exp5, Exp6}]. However, theoretical attempts to understand its microscopic reasons were restricted to the region of weak levitation, $\omega_c\tau \gtrsim 1$,  where delocalized states depart from the centers of Landau bands only slightly.

A minimal microscopic model which describes the levitation phenomenon, was proposed recently in Ref. [\onlinecite{MKR}] in terms of a certain random network. This model, named the $p$-$q$ model, captures the highly nontrivial interplay and competition of the disorder-induced scattering and the magnetic-field-induced weak orbital bending of electron trajectories, which leads to criticality. The mechanism governing the critical behavior in weak magnetic fields is very different from the one in strong fields, and leads to an even number (at least two) of channels on each link of the network, for two counterpropagating electron trajectories. Numerical and semi-analytical analysis of Ref. [\onlinecite{MKR}] revealed the phase diagram of the $p$-$q$ model comprising an Anderson insulator and two quantum Hall phases, separated by a delocalization boundary, in agreement with predictions of the scaling theory. However, the semi-analytical treatment in Ref. [\onlinecite{MKR}] was carried out in the ``classical'' limit of strong disorder and was based on percolative arguments. It is by no means rigorous, and a more controlled analytical treatment of the $p$-$q$ model is desirable.

In this paper, we study the $p$-$q$ network model analytically. We use the SUSY approach of Refs. [\onlinecite{GruzReadSach}, \onlinecite{GruzLudRead}] to map the $p$-$q$ model to an interacting superspin model. The superspins, which are certain irreducible representations of the Lie superalgebra u$(n,n|2n)$ ($n \geqslant 1$ is an integer), reside on the sites of a two-leg ladder. We treat the resulting superspin model by further mapping it to a supersymmetric nonlinear sigma model. Prior to the sigma model treatment, in order to develop intuition, we discuss the case of the spin ladder where the superspins are replaced by the more familiar su$(2)$ spins. The motivation to do this is the following. In the SUSY formalism, the single CC network model is mapped to a single superspin chain, and the phenomenology of the IQH transition is that of the opening a gap in the superspin spectrum due to dimerization of the bonds of the chain. This is the same phenomenology that governs the critical behavior of the usual su$(2)$ spin chain with dimerization, even though the numerical values of critical exponents are different. We believe that in our case of the superspin ladder, we can still determine the overall structure of the phase diagram by considering ordinary su$(2)$ staggered spin ladders. These were extensively studied [\onlinecite{ShNT, MDShS, Sierra, WN, MDDS, NYI}], and we use results of this research to gain intuition into the phase diagram of the $p$-$q$ model in the regions inaccessible to the sigma model description. Utilizing the developed machinery, we study localization properties of three additional random network models related to the $p$-$q$, and discuss proposals for their physical applications.

The paper is organized as follows. In Sec. \ref{sec:ladder}, we describe the $p$-$q$ model and map its anisotropic version to a superspin ladder. In Sec. \ref{sec:su2-ladder}, we discuss the su$(2)$ $\text{spin-}\frac{1}{2}$ counterpart of the superspin ladder and describe its quantum phase diagram. We introduce coherent states and develop a sigma model description of the model by formulating path integral over these states in Sec. \ref{sec:coherent-states}. We investigate three additional random network models, related to the $p$-$q$ model, and discuss their physical implications in Sec. \ref{sec:related-models}. Our results are summarized in the last Section. Appendices contain technical details of some of the derivations.

\section{The $p$-$q$ model and its mapping to a superspin ladder}
\label{sec:ladder}

%%%%%%%%%%%%%%%%%%%%%%%%%%%%%%%%%%%%%%%%%%%%%
\begin{figure}[t]
%\hfill
\includegraphics[width=0.9\columnwidth]{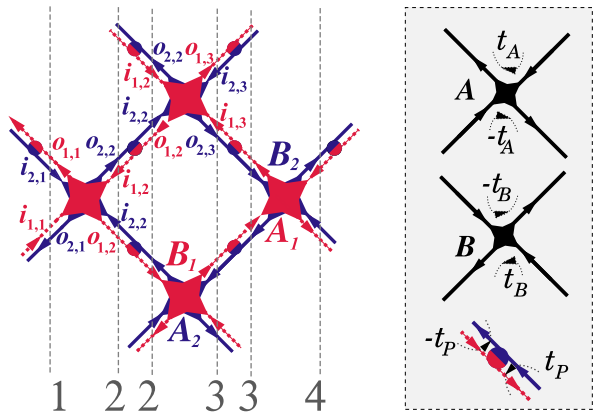} \vskip 2mm
\caption{(Color online) Left: the $p$-$q$ model.  The red and blue links and nodes form two copies of the CC network model. Vertical dashed lines denote sites of the spin ladder obtained from the network model in an anisotropic limit. Right: elementary scattering events and the corresponding amplitudes.}
\label{fig:Network}
\end{figure}
%%%%%%%%%%%%%%%%%%%%%%%%%%%%%%%%%%%%%%%%%%%%%%%%%%%%%

The simplest version of the $p$-$q$ network we study in this paper is depicted in Fig. \ref{fig:Network}. The network consists of two counterpropagating subnetworks, shown in blue and red in the figure. We label objects and quantities related to the two subnetworks by subscripts $l$ taking values $1$ for the red subnetwork and $2$ for the blue subnetwork. The nodes on each subnetwork are of two kinds: $A_1$ and $B_1$ on the red one, and $A_2$ and $B_2$ on the blue one. Vertical columns of links (shown by dashed lines in Fig. \ref{fig:Network}) are labeled by an integer subscript $n$ which will play the role of the discrete space index upon mapping to the superspin ladder later on. On each link the two counterpropagating channels carry fluxes $(i_{1,n}, o_{2,n})$ or $(o_{1,n}, i_{2,n})$, where $i$ stands for ``incoming'' and $o$ for ``outgoing'' (relative to a particular node) fluxes. The fluxes mix when propagating along a link, and the mixing is described by a $2\times2$ link scattering matrix $S(t_P)$:
\begin{align}
\label{pMatrices}
\begin{pmatrix} o_{1,n} \\ o_{2,n} \end{pmatrix} &= S(t_P) \begin{pmatrix} i_{1,n} \\ i_{2,n} \end{pmatrix}.
\end{align}
Scattering on links is illustrated by two-colored (red and blue) circles in Fig. \ref{fig:Network}. These circles separate each link into two ``half-links'' (each carrying the same ``spatial'' index $n$ as the whole link).

At the nodes of the network the scattering of fluxes is described by $4 \times 4$ node scattering matrices:
\begin{align}
\label{qMatrices}
\begin{pmatrix} o_{1,1} \\o_{1,2} \\ o_{2,1} \\ o_{2,2} \end{pmatrix} &=
\begin{pmatrix} S(t_{B_1}) & \block{cc}{0 & 0 \\ 0 & 0} \\
\block{cc}{0 & 0 \\ 0 & 0} & S(t_{A_2})\end{pmatrix}
\begin{pmatrix} i_{1,1} \\i_{1,2} \\ i_{2,1} \\ i_{2,2} \end{pmatrix}, \nonumber \\
\begin{pmatrix} o_{1,2} \\ o_{1,3} \\ o_{2,2} \\ o_{2,3}  \end{pmatrix} &=
\begin{pmatrix} S(t_{A_1}) & \block{cc}{0 & 0 \\ 0 & 0} \\
\block{cc}{0 & 0 \\ 0 & 0} & S(t_{B_2}) \end{pmatrix}
\begin{pmatrix} i_{1,2} \\i_{1,3} \\ i_{2,2} \\ i_{2,3} \end{pmatrix}.
\end{align}
Notice that scattering at the nodes does not mix the two subnetworks, and we illustrate the nodal scattering matrices by the red and blue star-shaped regions in Fig. \ref{fig:Network}. Each subnetwork is a copy of the CC network, and they are coupled by the link scattering. The $2\times 2$ scattering matrices in Eq. (\ref{qMatrices}) are chosen as
\begin{align}
S(t) =  \begin{pmatrix} \sqrt{1-t^2} & t \\ - t & \sqrt{1-t^2} \end{pmatrix},
\end{align}
with non-random parameters $t_{A_1}$, etc.

Disorder in the model is introduced through random phases acquired by fluxes along the half-links, that is, between the scattering events on the links and at the nodes. The total scattering matrix for propagation along a link in Eq. (\ref{pMatrices}) has the form
\begin{align}
\label{Sdecomposition}
S(t_P) = \begin{pmatrix} e^{i \phi_1} & 0 \\ 0 & e^{i \phi_2} \end{pmatrix}
\begin{pmatrix} \sqrt{1-t_P^2} & t_P \\ - t_P & \sqrt{1-t_P^2} \end{pmatrix}
\begin{pmatrix} e^{i \phi_3} & 0 \\ 0 & e^{i \phi_4} \end{pmatrix}.
\end{align}
The phases $\phi_i$ are uniformly distributed on the interval $[0, 2\pi)$, and are completely analogous to the random phases in the CC model.

The network model described above is a generalization of the original $p$-$q$ model [\onlinecite{MKR}] which contained only two parameters: $p$, the probability of backscattering on links, and $q$, quantifying the the asymmetry between scattering probabilities to the left and to the right for any incident channel. Physically, parameter $p$ is  related to the local (Drude) conductivity, and the quantity
\begin{align}
\gamma \equiv 2q - 1
\label{gamma}
\end{align}
characterizes the strength of a weak non-quantizing magnetic field. Our model reduces to the original one if we specify
\begin{align}
t_P^2 &= p, & t_{A_1}^2 &= t_{A_2}^2 = q, & t_{B_1}^2 & = t_{B_2}^2 = 1-q.
\label{eq:original-p-q}
\end{align}
This choice of scattering parameters satisfies the isotropy conditions $t_{A_l}^2 + t_{B_l}^2 = 1$ $(l = 1, 2)$. It is useful to relax these conditions and consider more general networks, where $t_{A_1} \neq t_{A_2}$, $t_{B_1} \neq t_{B_2}$, as well as $t_{A_l}^2 + t_{B_l}^2 \neq 1$, and we will do this in the following.

In this paper, we treat the (generalized) $p$-$q$ model analytically, utilizing the SUSY approach [\onlinecite{GruzReadSach}, \onlinecite{GruzLudRead}]. To this end we regard the vertical direction in Fig. \ref{fig:Network} as the (imaginary) time $\tau$. We denote the elementary time interval, the vertical separation between the middle points of two adjacent half-links, by $a_{\tau}$. We introduce bosons and fermions on each of the channels on each half-link (one species per advanced/retarded sectors). One channel on each half-link goes ``up'' (along the time direction) and another goes ``down''. The difference in the direction of the two channels leads to different commutation relations for creation and annihilation operators of the ``up'' and ``down'' particles. The time evolution operator $U$ describes the dynamics of bosons and fermions in the discrete imaginary time $\tau$. It possesses SUSY in the sense that it commutes with generators of the Lie superalgebra u$(1,1|2)$ for any realization of random phases. The SUSY form of $U$ can be explicitly averaged over the random phases on the half-links. This procedure results in the projection onto certain mutually dual irreducible representations (irreps) $R$ and $\bar R$ of u$(1,1|2)$ for each half-link [\onlinecite{GruzReadSach}, \onlinecite{GruzLudRead}] (see Appendix \ref{sec:appendix-SUSY-1}).

An explicit form of $U$ in the second quantized formulation can be written as an exponential of a quadratic form in creation and annihilation operators (see details in Appendix \ref{sec:appendix-SUSY-2}). Here, we will only be concerned with its time-continuous (Hamiltonian) version. In order to achieve the Hamiltonian description, we consider a general anisotropic version of the $p$-$q$ network with parametrization
\begin{align}
t_P^2 &= p, & t_{A_l}^2 &= \epsilon q_l, & t_{B_l}^2 &= \epsilon (1-q_l).
\label{eq:anisotropic-p-q}
\end{align}
At $\epsilon = 1$ isotropy is restored, and when $q_1 = q_2 = q$, the original $p$-$q$ model is recovered [see Eq. (\ref{eq:original-p-q})]. We now proceed by taking the limit $t_P, \epsilon \ll 1$, which introduces strong anisotropy in the network but should not affect critical properties of the system [\onlinecite{footnote}]. In this anisotropic limit, each subnetwork can be viewed as a collection of vertical zigzag channels with alternating (up or down) overall direction of fluxes. A weak hopping between adjacent channels takes place when they come close at the ($4\times4$) nodes, given by the amplitudes $t_{A_1}, t_{B_1}$ and $t_{A_2}, t_{B_2}$. In addition, the two subnetworks are weakly coupled by scattering on the links, with the hybridization amplitude $t_P \ll 1$.

Upon taking the $\tau$-continuum limit (see Appendix \ref{sec:appendix-SUSY-3} for details), the resulting disorder-averaged evolution operator $[U]$ can be written as
\begin{align}
[U] = \exp{\Big( - \int_0^\beta \! d\tau\, \mathcal{H} \Big)},
\end{align}
where the one-dimensional Hamiltonian $\mathcal{H}$ is
\begin{align}
\label{Hamiltonian}
\mathcal{H} &= - \str\! \sum_k \Big[J_{\perp} \Big( \mathcal{S}^{(1)}_{2k-1} \bar{\mathcal{S}}^{(2)}_{2k-1}
+ \bar{\mathcal{S}}^{(1)}_{2k} \mathcal{S}^{(2)}_{2k}\Big) \nonumber \\
& + J\Big((1+\gamma_1) \mathcal{S}^{(1)}_{2k-1}\bar{\mathcal{S}}^{(1)}_{2k}
+ (1-\gamma_1) \bar{\mathcal{S}}^{(1)}_{2k} \mathcal{S}^{(1)}_{2k+1}\Big) \nonumber   \\
& + J\Big((1-\gamma_2)\bar{\mathcal{S}}^{(2)}_{2k-1} \mathcal{S}^{(2)}_{2k}
+ (1+\gamma_2)\mathcal{S}^{(2)}_{2k} \bar{\mathcal{S}}^{(2)}_{2k+1}\Big) \Big].
\end{align}
The notation we use here is the following. The superspins $\mathcal{S}$ and $\bar{\mathcal{S}}$ are graded matrices whose matrix elements are the generators of u$(1,1|2)$ acting in the irreps $R$ and $\bar R$ (see Appendix \ref{sec:appendix-SUSY-1} for details). The superspins are labeled by two indices according to their position on the sites of a two-leg ladder: the superscript $l$ denotes a leg of the ladder (which corresponds to the subnetwork $l$), and the subscript refers to the position along the ladder. The parameters $J$, $J_\perp$, and $\gamma_l$ are related to the scattering amplitudes $p$ and $q_l$ and the anisotropy parameter $\epsilon$ by
\begin{align}
J &= \frac{t_{B_l}^2 + t_{A_l}^2}{2} = \frac{\epsilon}{2}, & J_\perp &=2t_P^2 = 2p, \nonumber \\
\gamma_l &= 2q_l - 1 = \frac{t_{A_l}^2 - t_{B_l}^2}{t_{B_l}^2 + t_{A_l}^2}.
\end{align}

In the original $p$-$q$ model, there are only two parameters, $p$ and $q$. The corresponding superspin chain is obtained by making the dimerization parameters $\gamma_l$ equal:
\begin{align}
\gamma_1 = \gamma_2 = \gamma.
\label{staggering}
\end{align}
In this case, the superspin Hamiltonian is
\begin{align}
\label{Hamiltonian-p-q}
\mathcal{H}_{p\text{-}q} &= - \str\! \sum_k \Big[J_{\perp} \Big( \mathcal{S}^{(1)}_{2k-1} \bar{\mathcal{S}}^{(2)}_{2k-1}
+ \bar{\mathcal{S}}^{(1)}_{2k} \mathcal{S}^{(2)}_{2k}\Big) \nonumber \\
& + J\Big((1+\gamma) \mathcal{S}^{(1)}_{2k-1}\bar{\mathcal{S}}^{(1)}_{2k}
+ (1-\gamma) \bar{\mathcal{S}}^{(1)}_{2k} \mathcal{S}^{(1)}_{2k+1}\Big) \nonumber   \\
& + J\Big((1-\gamma)\bar{\mathcal{S}}^{(2)}_{2k-1} \mathcal{S}^{(2)}_{2k}
+ (1+\gamma)\mathcal{S}^{(2)}_{2k} \bar{\mathcal{S}}^{(2)}_{2k+1}\Big) \Big].
\end{align}

\begin{figure}[t]
\centerline{\includegraphics[width=0.9\columnwidth]{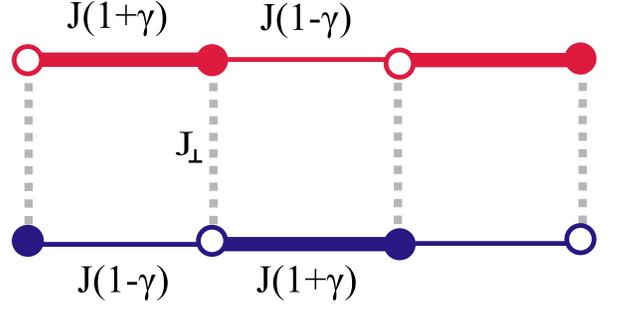}}
\caption{(Color online) The superspin ladder described by the Hamiltonian $\mathcal{H}_{p\text{-}q}$ [Eq. (\ref{Hamiltonian-p-q})]. The empty (filled) circles on the sites of the ladder indicate the superspins $\cal S$ ($\bar{\mathcal S}$) in the $R$ ($\bar{R}$) representation.}
\label{fig:pq-ladder}
\end{figure}

The Hamiltonians (\ref{Hamiltonian}) and (\ref{Hamiltonian-p-q}) describe the dynamics of superspins arranged on sites of a two-leg ladder, with interactions between nearest neighbors. The two legs of the ladder correspond to the two subnetworks that are colored by red and blue in Fig. \ref{fig:Network}. We keep the same colors to illustrate the arrangement of the superspins $\cal S$ (filled circles) and $\bar{\mathcal{S}}$ (empty circles) on the ladder in Fig. \ref{fig:pq-ladder}. The structure of the original network model (specifically, the counter-propagation of fluxes on the links) results in the fact that superspins $\cal S$ and $\bar{\mathcal{S}}$ (or irreps $R$ and $\bar R$) alternate both along the legs and along the rungs. The superspin exchange has the u$(1,1|2)$-invariant bilinear form, $\str\mathcal{S}\bar{\mathcal S}$. This form is diagonalized on each bond by decomposing the tensor product $R \otimes {\bar R}$ into irreps. This decomposition necessarily contains a singlet of u$(1,1|2)$, since two dual representations are involved. With the above signs and positive $J$ and $J_{\perp}$, the pair exchange energy is minimized in the singlet state on a bond, so that all the couplings in Eqs. (\ref{Hamiltonian}) and (\ref{Hamiltonian-p-q}) are {\it antiferromagnetic}.

Our goal is to analyze the superspin ladder described by the Hamiltonian (\ref{Hamiltonian-p-q}) by mapping it to a supersymmetric sigma model, and determining its phase diagram. This will be achieved in Sec. \ref{sec:coherent-states}. Meanwhile, to gain some intuition, in the next section we will consider the analog of Eq. (\ref{Hamiltonian-p-q}) for the ladder where the u$(1,1|2)$ superspins are replaced by the usual su$(2)$ spins $S=\frac{1}{2}$. Although this replacement alters quantitative characteristics of the phase diagram and the phase transitions of the spin ladder, we believe that the qualitative nature of the phase diagram remains intact.

\section{su$(2)$ spin ladder}
\label{sec:su2-ladder}

\begin{figure}[t]
\centerline{\includegraphics[width=\columnwidth]{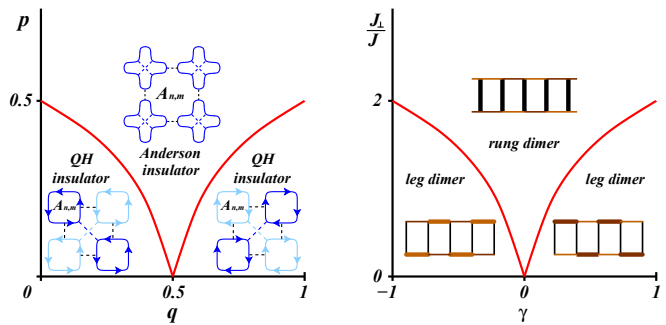}}
\caption{(Color online) Phase diagrams of the $p$-$q$ model from Ref. [\onlinecite{MKR}] (left) and the su$(2)$ staggered spin ladder from Ref. [\onlinecite{MDShS}] (right). Different massive phases of the staggered spin ladder are schematically illustrated by different thickness of bonds representing different local spins singlet (dimer) amplitudes.}
\label{PhaseDiag}
\end{figure}

In this section, we analyze the two-leg su$(2)$ spin ladder described by the following Hamiltonian:
\begin{align}
\label{SSL}
H &= J \sum\limits_{l,k} [1 + (-1)^{k+l}\gamma] \mathbf{S}_k^{(l)} \cdot \mathbf{S}_{k+1}^{(l)}
+ J_\perp \sum\limits_k \mathbf{S}_k^{(1)} \cdot \mathbf{S}_k^{(2)}.
\end{align}
Here, $l = 1,2$ labels the legs of the ladder, and $k$ labels the rungs, while $\mathbf{S}$ are the $S = \frac{1}{2}$ spins. This is exactly the Hamiltonian  studied by Mart\'{i}n-Delgado, Shankar, and Sierra [\onlinecite{MDShS}], who argued that its phase diagram includes three different massive phases separated by two critical lines (see the right part of Fig.~\ref{PhaseDiag}). Here we will give general qualitative arguments leading to this phase diagram, complemented by our bosonization analysis for the asymptotic behavior of the critical lines near $|\gamma|\sim 1$ and known results [\onlinecite{WN}, \onlinecite{MDDS}] for their behavior at $|\gamma| \ll 1$.

%%%%%%%%%%%%%%%%%%%%%%%%%%%%%%%
%%%%%%%%%%%%%%%%%%

The phase diagram of the spin ladder described by Eq. (\ref{SSL}) on the $(\gamma, J_\perp/J)$ plane has three distinguished points where the nature of the ground state is very transparent. The first two points are $\gamma =\pm 1$, $J_\perp/J = 2$, and they correspond to single homogeneous antiferromagnetic Heisenberg chains, with sites connected alternatively along the rungs and along the legs. These spin chains are critical. Keeping $\gamma = \pm 1$ but changing $J_\perp/J$ leads to dimerization of the chains and opens up a gap in the spectrum. The third point $\gamma=0$, $J_\perp/J=0$ corresponds to two decoupled homogeneous antiferromagnetic Heisenberg chains and is also critical. Keeping $\gamma = 0$ but turning $J_\perp > 0$ leads to a massive phase with dimers on the rungs. If instead we keep the chains decoupled ($J_\perp = 0$) but introduce dimerization in the couplings along the chains ($\gamma \neq 0$), then the chains also become massive, with dimers on one sublattice of each chain. The three critical points are connected by continuous critical lines which separate the massive phases, as shown in Fig.~\ref{PhaseDiag}.

Let us now focus on the vicinity of the points $\gamma =\pm 1$, $J_\perp/J = 2$. We relabel the sites along the ``snaking'' path on the ladder (this path represents the single critical spin chain). The precise relabeling depends on the sign of $\gamma$. Then, the Hamiltonian (\ref{SSL}) can be written as
\begin{align}
H &= H_0 + H_\epsilon + H_\eta,
\label{H-perturbations}
\end{align}
which represents the single homogeneous (critical) spin chain
\begin{align}
H_0=\frac{J_\perp + J(1+|\gamma|)}{2}\sum_k \mathbf{S}_k \cdot \mathbf{S}_{k+1}
\end{align}
subject to two weak perturbations
\begin{align}
H_\epsilon &= \epsilon \sum_k(-1)^k \mathbf{S}_k \cdot \mathbf{S}_{k+1}, & \epsilon &= \frac{J_\perp - J(1+|\gamma|)}{2}, \\
H_\eta &= \eta \sum_k \mathbf{S}_{2k} \cdot \mathbf{S}_{2k+3}, & \eta &= J(1-|\gamma|).
\end{align}

The perturbed Hamiltonian (\ref{H-perturbations}) can be studied utilizing the standard (Abelian) bosonization approach (see, for example, Ref. [\onlinecite{GNT-book}]). The low-energy physics is described in terms of the massless Bose field $\phi(x)$ and its conjugate momentum $\Pi(x)$. In terms of these fields the unperturbed Hamiltonian becomes quadratic (non-interacting):
\begin{align}
\label{2bf}
H_0 \rightarrow \frac{v_s}2 \int dx \left[\Pi^2(x)+\bigl(\partial_x\phi(x)\bigr)^2\right] + \cdots,
\end{align}
where
\begin{align}
v_s \sim \frac{1}{2} [J_\perp + J(1+|\gamma|)]a_0
\end{align}
is the spin wave velocity, $a_0$ is the lattice spacing, and the dots stand for irrelevant terms. In the same manner the perturbations in the bosonized form become
\begin{align}
H_\epsilon &\rightarrow h_\epsilon \int dx \cos\sqrt{2\pi}\phi(x), & h_\epsilon &=  \frac{\lambda \epsilon}{a_0}, \\
H_\eta &\rightarrow h_\eta\int dx \cos\sqrt{2\pi}\phi(x), & h_\eta &=  \frac{\lambda \eta}{6 a_0},
\end{align}
where $\lambda$ is a non-universal parameter of order one. We see that both perturbations have the same form, and are strongly relevant in the renormalization group (RG) sense. Hence, the system is critical if the overall coefficient $h_\epsilon + h_\eta$, vanishes:
\begin{align*}
h_\epsilon+h_\eta=\frac\lambda{a_0} \left[\frac{J_\perp - J(1+|\gamma|)}2+\frac J6(1-|\gamma|)\right] = 0.
\end{align*}
Thus, in terms of the small parameter $(1-|\gamma|)$, the critical line near $\gamma = \pm1$, $J_\perp/J = 2$ is given by the equation
\begin{align}\label{critline1}
\frac{J_\perp}J = 2- \frac43(1-|\gamma|),
\end{align}
which is in an excellent agreement with known numerical results [\onlinecite{NYI}].

In the vicinity of the third point, $\gamma=0$, $J_\perp/J=0$, the situation is more complicated. A consistent non-perturbative description of the critical behavior near this point was developed in Ref. [\onlinecite{WN}] by mapping the lattice model onto an $\text{O}(3)\times Z_2$-symmetric theory of four massive Majorana fermions. The resulting behavior of the critical line,
\begin{align}
\label{critline2}
\frac{J_\perp}J \propto |\gamma|^{2/3},
\end{align}
was previously predicted from heuristic arguments [\onlinecite{MDDS}], and subsequently confirmed by numerical simulations [\onlinecite{NYI}]. The crossover exponent
\begin{align}
\phi = \frac{2}{3} = \frac{y_\perp}{y_\gamma}
\label{crossover-exp}
\end{align}
that appears in Eq. (\ref{critline2}) is the ratio of the RG eigenvalues
\begin{align}
y_\perp &= 1, & y_\gamma & = 3/2,
\label{RG-eignvalues}
\end{align}
which are related in the usual way to the dimensions $x_\perp = 1$, $x_\gamma = 1/2$ of the two relevant operators (coupling of the two spin chains and the dimerization) near the point of decoupled critical spin chains.

Thus, the phase diagram of the spin-$\frac{1}{2}$ Hamiltonian Eq.~(\ref{SSL}) has the form shown on the right panel in Fig.~\ref{PhaseDiag} with asymptotic behaviors near the end-points given by Eqs. (\ref{critline1}) and (\ref{critline2}). Notice that the simple relation $J_\perp/J=2|\gamma|^{2/3}$ has the above asymptotic forms and represents an approximate analytic form of the critical line reasonably close to the numerically determined one [\onlinecite{NYI}].

As we have mentioned in the Introduction, we believe that it is quite reasonable to expect that the original superspin ladder Eq.~(\ref{Hamiltonian}) has a similar phase diagram, though with asymptotic behaviors different from Eqs. (\ref{critline1}) and (\ref{critline2}). Similar to the analogy between the IQH transition in strong fields and a single spin chain, we have now the analogy between the low-field quantum Hall transition and the quantum criticality of the staggered spin ladder Eq. (\ref{SSL}). Within this analogy, the Anderson insulator and the two different quantum Hall insulating phases of the $p$-$q$ model correspond to the rung-dimer and two different leg-dimer phases of the staggered spin ladder (see Fig.~\ref{PhaseDiag}).

\section{Coherent states and the nonlinear sigma model}
\label{sec:coherent-states}

In this section, we return to the study of the staggered superspin ladder described by the Hamiltonian (\ref{Hamiltonian}). To obtain the structure of its phase diagram, we map it to a supersymmetric sigma model.

It is advantageous to generalize the $p$-$q$ network model to a variant with an arbitrary number $N$ of channels propagating in both directions on each link. The scattering matrices on the links become
\begin{align}
S(t_P) =
\begin{pmatrix}
U_1 & 0 \\ 0 & U_2
\end{pmatrix}
\begin{pmatrix}
\sqrt{1 - T_P^2} & T_P \\ - T_P & \sqrt{1 - T_P^2}
\end{pmatrix}
\begin{pmatrix}
U_3 & 0 \\ 0 & U_4
\end{pmatrix},
\label{polar}
\end{align}
with matrices $U_i \in \text{U}(N)$ describing the mixing of the fluxes on the half-links. These matrices are analogs of the random phases in the simpler case of Eq. (\ref{Sdecomposition}), and are taken to be uniformly distributed  over the U($N$) group. The orthogonal matrix in the middle, describing the  scattering of fluxes at a node, is parametrized by the diagonal in the channel index $N \times N$ matrix
\begin{align}
T_P = \diag (t_P,\ldots, t_P).
\label{T}
\end{align}
The number of channels $N$ will play the role of a large parameter that will control the gradient expansion in the derivation of the sigma model.

We further introduce $n$ copies (replicas) of bosons and fermions associated with every channel. This allows us to consider multi-point correlation functions, or higher moments of Green's functions. In this situation, the supersymmetry of the model becomes u$(n,n|2n)$, and the superspins $\mathcal{S}$ associated with the up-going links are now certain highest-weight irreps $R_N$ of u$(n,n|2n)$. Similarly, for the down-going links we obtain the superspins $\bar{\mathcal{S}}$ which are lowest weight irreps ${\bar R}_N$ of u$(n,n|2n)$ (dual to $R_N$) (see Appendix \ref{sec:appendix-coherent-states} for details).

The next step is to introduce the superspin coherent states [\onlinecite{Perelomov}] and the functional integral over them. This step is quite analogous to the coherent-state path-integral derivation in Refs. [\onlinecite{Wiegmann}, \onlinecite{Read+Sachdev}], so we relegate details to Appendix \ref{sec:appendix-coherent-states}. As a result, a single superspin $\cal S$ with the Hamiltonian $\cal H [\cal S]$ is described by the following imaginary-time path integral:
\begin{align}
Z = \int {\cal D} \Omega \exp \Big(- \int_0^\beta \! d\tau \, {\cal L} \Big),
\end{align}
where the Lagrangian $\cal L$ is:
\begin{align}
\label{RCohStateAct}
{\cal L} &= {\cal L}_B[\Omega] + \mathcal{H}[N\Omega(\tau)/2],  \nonumber \\
{\cal L}_B[\Omega] &=  \frac{N}{4}\int_0^1 \! du \str [\Omega(\tau, u) \partial_u
\Omega(\tau, u) \partial_\tau \Omega(\tau,u )].
\end{align}
Here $\Omega(\tau)$ is a periodic matrix function, $\Omega(L_\tau) = \Omega(0)$, taking values in the super-coset space $G/H$, where $G = \text{U}(n,n|2n)$ and $H = \text{U}(n|n) \times \text{U}(n|n)$. The matrix
\begin{align}
\Lambda = \diag (I_{2n}, -I_{2n})
\label{}
\end{align}
plays the role of the origin in $G/H$, and other points in this space can be reached from the origin by the adjoint action of elements $g \in G$:
\begin{align}
\Omega = g \Lambda g^{-1}.
\end{align}
Also, $\Omega (\tau, u)$ is a homotopy between $\Lambda$ and $\Omega(\tau)$ as $u$ goes from $0$ to $1$. For a single conjugate superspin $\bar{\mathcal{S}}$ with Hamiltonian $\mathcal{H}(\bar{\mathcal{S}})$ the Lagrangian is
\begin{align}
\label{barRCohStateAct}
{\cal L} &= -{\cal L}_B[\Omega] + \mathcal{H}[-N\Omega(\tau)/2],
\end{align}
which can formally be obtained from Eq. (\ref{RCohStateAct}) by flipping the sign of all matrices $\Omega$ since, in our notation,
\begin{align}
{\cal L}_B[-\Omega] =  - {\cal L}_B[\Omega].
\label{Berry-term-odd}
\end{align}

In taking the continuum limit to obtain a nonlinear sigma model, we closely follow a similar derivation for SU$(N)$ antiferromagnets [\onlinecite{ArovAuer}, \onlinecite{Read+Sachdev}]. First, we define the fields $\Omega_k^{(l)}$ on each site of both legs ($l = 1, 2$) of the ladder. In terms of these fields, the Lagrangian for the superspin ladder with the Hamiltonian (\ref{Hamiltonian}) is
\begin{align}
{\cal L} &= \sum_{l,k} (-1)^{k+l} {\cal L}_B[\Omega_k^{(l)}] + \frac{J_\perp N^2}{4} \str \sum_k  \Omega_k^{(1)} \Omega_k^{(2)} \nonumber \\
&\quad + \frac{J N^2}{4} \str \sum_{l,k}\big[1 + (-1)^{k+l} \gamma_l \big] \Omega_k^{(l)} \Omega_{k+1}^{(l)}.
\label{Lagrangian}
\end{align}

In view of the staggered arrangement of the superspins on the ladder (the antiferromagnetic nature of the couplings), the expected ground state of the ladder is of N\'eel type. In the large-$N$ limit, we can treat the superspin ladder semiclassically and decompose the field $\Omega$ into the staggered ($Q$) and uniform ($L$) components, as well as their harmonics in the transverse direction (along the rungs of the ladder) $R$ and $M$:
\begin{align}
\Omega_{2k-1}^{(l)} &= Q_k - a[ (-1)^{l} (R_k - L_k) + M_k], \nonumber \\
\Omega_{2k}^{(l)} &= Q_k - a [(-1)^{l} (R_k + L_k) - M_k].
\label{Sierra-decomposition}
\end{align}
This decomposition is similar to the one used in Ref. [\onlinecite{Sierra}] (where the author used a Hamiltonian operator formalism). Both $Q$ and $L$ fields are taken to be the same on both legs of the ladder and are expected to be smoothly varying along the ladder. In the continuum limit all the fields except $Q$ happen to be massive and can be integrated out (see details in Appendix \ref{sec:appendix-sigma model}). It is worth mentioning here that the mass of the $R$ field is proportional to $J_\perp$, so we may expect the decomposition (\ref{Sierra-decomposition}) to become less and less meaningful and useful as we approach the point of decoupled chains $J_\perp = 0$. On the other hand, the masses of $M$ and $L$ are set by $J$, so they remain finite even at $J_\perp = 0$.

Integrating out all the fields except $Q$, we get the following nonlinear sigma model action:
\begin{align}
S = \frac{1}{8} \int \!\! \big[-\sigma_{xx} (\partial_x Q)^2 + 2 \sigma_{xy}  Q\partial_\tau Q \partial_x Q \big].
\label{NLSMAction}
\end{align}
Here and below we adopt the shorthand notation
\begin{align}
\int \equiv \int \!\! d\tau dx \, \str.
\label{shorthand}
\end{align}
The coefficients in the action (\ref{NLSMAction}) play the role of the bare dimensionless conductivities and are given by
\begin{align}
\sigma_{xx} &= \frac{N}{\mu} \sqrt{\mu(1 - \gamma_-^2) - \gamma_+^2}, & \sigma_{xy} &=  -\frac{N \gamma_+}{\mu},
\label{conductivities-two-leg-ladder}
\end{align}
where
\begin{align}
\gamma_\pm &= \frac{\gamma_1 \pm \gamma_2}{2}, & \mu &= 1 + \frac{J_\perp}{2J}.
\end{align}
The Hall conductivity is related to the so called $\theta$ angle by
\begin{align}
\theta &= 2\pi \sigma_{xy} = -\frac{2 \pi N \gamma_+}{\mu}.
\label{theta-angle}
\end{align}

For the spin ladder (\ref{Hamiltonian-p-q}), the expressions for the conductivities and the $\theta$ angle simplify:
\begin{align}
\sigma_{xx} &=
N \frac{\sqrt{1 - \gamma^2 + J_\perp/2J}}{1 + J_\perp/2J}, &
\sigma_{xy} &= -\frac{N\gamma}{1 + J_\perp/2J},
\label{sigma-xy-staggered-two-legs} \\
\theta &= -\frac{2N\pi\gamma}{1 + J_\perp/2J}.
\label{tt}
\end{align}

The sigma model (\ref{NLSMAction}) is critical when the $\theta$ angle is an odd multiple of $\pi$. Let us explore the single channel case $N=1$ in more details. In this case, the $\theta$ angle [Eq. (\ref{tt})] exactly coincides with the one derived in Ref.~[\onlinecite{MDShS}] for the two-leg $S=\frac{1}{2}$ staggered spin chain discussed in the previous section. The shape of the critical line $\theta = \pm \pi$ translates into the dependence
\begin{align}
\frac{J_\perp}{J} = 2 - 4(1 - |\gamma|).
\label{critical-lines-sigma model}
\end{align}
This result strongly deviates from more accurate results Eqs. (\ref{critline1}) and (\ref{critline2}). Still, we see that near $\gamma = \pm 1$, $J_\perp/J = 2$, this result agrees with Fig.~\ref{PhaseDiag} and Eq.~(\ref{critline1}) qualitatively. On the other hand, for $J_\perp/J \ll 1$, it yields two critical points at $\gamma = \pm \frac{1}{2}$, far from $\gamma=0$. This strongly disagrees with the phase diagram Fig.~\ref{PhaseDiag}. Obviously, the $J_\perp/J=0$, $\gamma = 0$ point is  massless as the two superspin chains, without any dimerization, get decoupled, or in the network model picture, we have two critical CC networks. We clearly see that the sigma model (\ref{NLSMAction}) is not an adequate description of the supesrspin system (\ref{Hamiltonian-p-q}), in the vicinity of the point $J_\perp/J=0$, $\gamma = 0$. The reason is that the fields $R_k$ that we have integrated out become massless as $J_\perp \rightarrow 0$. In this limit, it is not the sigma model field $Q$ that becomes critical, but its combination with the other degrees of freedom.

To describe analytically the low-energy behavior of our system near the point $J_\perp/J=0$, $\gamma = 0$, we need to correctly identify the degrees of freedom that remain massive and the ones that are lower in energy and eventually become critical. We have seen in the previous section that for su$(2)$ spin-$\frac{1}{2}$ case (with $N$=1) this was done in Ref. [\onlinecite{WN}], by mapping the system onto four weakly coupled Ising models. As we don't have a luxury to relate our superspin model to simpler systems such as the Ising model, we choose a different way. Namely, we make an alternative parametrization of the system by introducing two independent sets of $Q$- and $L$-fields for each leg by
\begin{align}
\Omega^{(l)}_{2k-1} &= Q^{(l)}_k - a L^{(l)}_k, &\Omega^{(l)}_{2k} &= Q^{(l)}_k + a L^{(l)}_k.
\end{align}
Making the dimerization parameters $\gamma_l$ (possibly) different on each leg, again, and lowering the leg index $l$ for fields in the continuum, we obtain the action
\begin{align}
S &= \frac{N}{4} \int \!\! \sum_l (-1)^{l+1} L_l Q_l \partial_\tau Q_l \nonumber \\
& -\frac{J N^2 a}{4} \int \! \sum_l \Big[[1 + (-1)^l \gamma_l]\partial_x Q_l \big(\partial_x Q_l - 2 L_l\big) + 2 L_l^2 \Big] \nonumber \\
& -\frac{J_\perp N^2 a}{8} \int \!\! \Big[\frac{1}{a^2}\big(Q_1 - Q_2\big)^2 + \big(L_1 - L_2\big)^2 \Big].
\label{coupNLSM}
\end{align}

The last line in this equation couples the legs of the ladder. However, at $J_\perp = 0$ they get decoupled. Then, integrating out the $L$ fields, we obtain two decoupled sigma models, one for each leg. These sigma models are critical only if $\gamma_1=0$ and $\gamma_2=0$. Thus, we see that the description (\ref{coupNLSM}) does capture the low-energy critical properties of our superspin system for $J_\perp = 0$, in contrast with the previous treatment through a single sigma model (\ref{NLSMAction}).

In the vicinity of the point $J_\perp/J=0$, $\gamma = 0$ the following arguments seem to be plausible (compare with the discussion of a similar action in Ref. [\onlinecite{OGM}], Sec. IIC). For a positive $J_\perp/J$, the first term in the second line of Eq. (\ref{coupNLSM}) is a relevant perturbation, leading in the infrared to ``locking'' of the fields
\begin{align}
Q_1 = Q_2 = Q.
\end{align}
Once the locking happens, the $L$ fields can be integrated out, which gives again the sigma model action (\ref{NLSMAction}) with the couplings (\ref{conductivities-two-leg-ladder}). When the dimerization parameters $\gamma_l$ are equal as in Eq. (\ref{staggering}), we recover the conductivities (\ref{sigma-xy-staggered-two-legs}) and the critical lines (\ref{critline1}). We conclude that for $J_\perp > 0$, the critical point at $J_\perp/J=0$, $\gamma = 0$, splits into two critical lines.

We thus believe that the critical lines on the phase diagram of the superspin ladder (\ref{Hamiltonian-p-q}) include the $J_\perp/J=0$, $\gamma = 0$ point, and close to it have the form
\begin{align}
\frac{J_\perp}{J} \propto |\gamma|^\phi,
\end{align}
shown (qualitatively) by the curve in Fig. \ref{PhaseDiag}. The analytical value of the crossover exponent $\phi$ in this case is not known, since, unlike in the su(2) case, we do not know the dimensions $x_\gamma$ and $x_\perp$ of the relevant operators or the values of the corresponding RG eigenvalues $y_\gamma$ and $y_\perp$. However, $y_\gamma = 1/\nu$, where $\nu \approx 2.6$ is the localization length exponent known numerically [\onlinecite{}. This gives $\phi = \nu y_\perp \approx 2.6 y_\perp$. If we assume that $y_\perp \sim O(1)$ (not too small, similar to the case of the su(2) spin chain), we obtain that $\phi > 1$. This is in agreement with the results for the form of the phase diagram established semiclassically in Refs.  [\onlinecite{MKR}, \onlinecite{WeOrt}]. The second of these papers reports $\phi \approx 2$. Notice also that the original results of Khmelnitskii [\onlinecite{Khmelnitskii}] and Laughlin [\onlinecite{Laughlin}] combined with (non-rigorous) heuristic arguments translate into the linear dependence, $\phi =1$ (see Ref. [\onlinecite{WeOrt}]).

\section{Random network models related to the $p$-$q$ model}
\label{sec:related-models}

While our primary goal was to use the supersymmetry method to study the phase diagram of the $p$-$q$ model, we can apply this method to other random network models. In this section, we describe three random network models related to the $p$-$q$ model, and study their localization properties using the method developed in previous sections. We also discuss their physical implications in the subsequent section.

As we have seen above, the $p$-$q$ model can be thought of as two CC models (two layers) coupled by the $p$ scattering in the middle of each link. In this case, the pairs of fluxes mixed on the links are counterpropagating, and phases acquired by each flux between any two scattering events are random and independent. Also, parameters of the nodes of each CC subnetwork of the $p$-$q$ model are chosen in the way that any flux incoming to a node scatters to the left with the same probability, equal to $q$, for all the nodes.

\begin{figure*}[btp]
\centerline{\includegraphics[width=\textwidth,angle=0,clip]{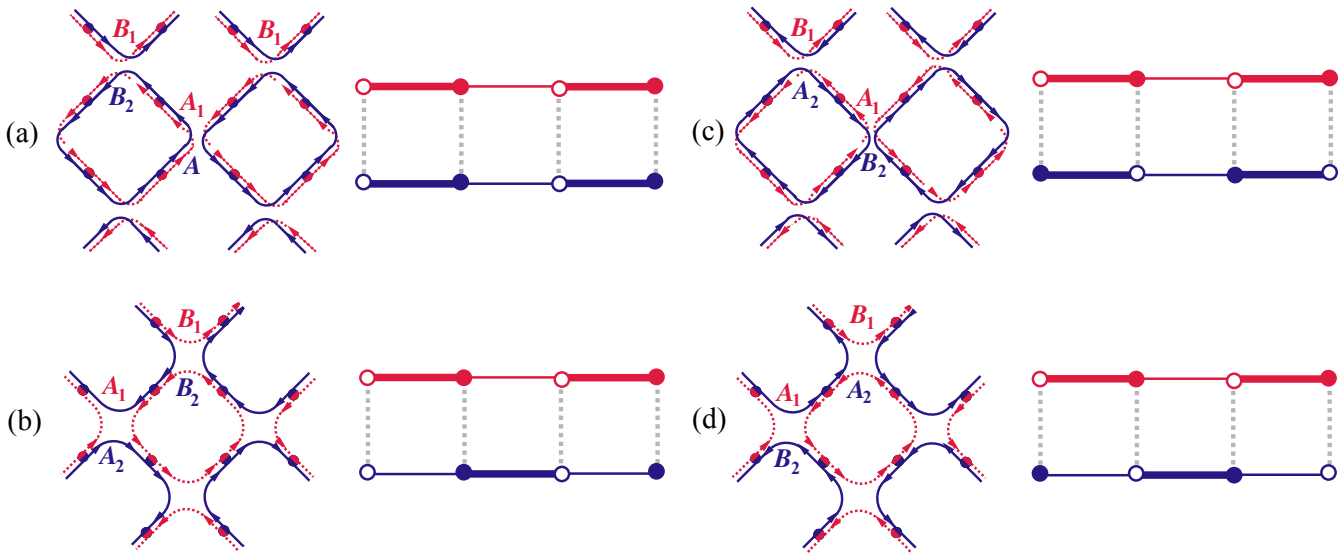}}
\caption{(Color online) Two-channel networks related to the $p$-$q$ model, and the corresponding spin ladders. In the network model, the lines indicate primary direction of scattering at the nodes. Strong scattering between adjacent sites in the network model leads to strong couplings in the spin chain.  The empty (filled) circles in the spin ladder indicate that the site has superspins in the $R$ ($\bar{R}$) representation.}
\label{diffnets}
\end{figure*}

Two CC networks can be coupled in other ways, so that the two fluxes on the resulting links are either copropagating or counterpropagating. At the nodes we also distinguish two possibilities, compatible with the requirement of overall isotropy: each incoming flux scatters to the left with the same probability, as is the case of the $p$-$q$ model, or the probability of scattering to the left for two of the incoming fluxes equals to the probability of scattering to the right for the other two incoming fluxes. Overall, we end up with four two-channel networks illustrated in Fig. \ref{diffnets}, one of which is the $p$-$q$ model shown in Fig. \ref{diffnets}(d).

In the next subsection we will demonstrate that, similar to the $p$-$q$ model, the phase diagram of the network Fig. \ref{diffnets}(a) includes three regions of localized states separated by lines of extended states. On the other hand, as we will demonstrate, in the network models shown in Figs. \ref{diffnets}(b) and \ref{diffnets}(c) all the states are localized.

%%%%%%%%%%%%%%%%%%%%%%%%%%%%%%%
%%%%%%%%%%%%%%%%%%

\subsection{Models and their phase diagrams}

We begin our consideration with the network with copropagating channels, illustrated in Fig. \ref{diffnets}(a). In notations of Sec. \ref{sec:ladder}, the scattering probabilities at the nodes are chosen exactly the same as in the $p$-$q$ model [see Eq. (\ref{eq:anisotropic-p-q})]. However, in the present case the nodes $A_1$ and $A_2$ are located on top of each other, as well as the nodes $B_1$ and $B_2$. As before, randomness is encoded in independent random phases acquired by each flux between scattering events at the nodes and on the links.

It is instructive to observe that the $p$-$q$ model transforms into this network upon a shift of one of the constituting CC networks by one lattice constant, along either of the two directions parallel to the links. Such a transformation makes two fluxes on the links copropagating, as well as brings the resulting nodes into the form illustrated in Fig.~\ref{diffnets}(a).

The copropagating fluxes on the links make this problem essentially different from the $p$-$q$ model. The second-quantized transfer matrices for the links are given in Ref. [\onlinecite{GruzReadSach}] (where they appeared in the study of the so-called ``chiral metal''). In the time continuum limit they lead to a {\it ferromagnetic} interaction between the superspins on the two legs of the superspin ladder. This means that the relevant terms in the Hamiltonian are now $- J_\perp \str\mathcal{S}^{(1)}_k{\mathcal S}^{(2)}_k$, and favor the representation with the largest highest-weight in the decomposition $R \otimes R$.

The node transfer matrices produce antiferromagnetic couplings along the legs of the ladder, as before. Thus, the disorder-averaged supersymmetric Hamiltonian associated with the network in Fig. \ref{diffnets}(a) is
\begin{align}
\mathcal{H}_a &= - \str\! \sum_k \Big[J_{\perp} \Big( \mathcal{S}^{(1)}_{2k-1} \mathcal{S}^{(2)}_{2k-1}
+ \bar{\mathcal{S}}^{(1)}_{2k} \bar{\mathcal{S}}^{(2)}_{2k}\Big) \nonumber \\
& + J\Big((1+\gamma_1) \mathcal{S}^{(1)}_{2k-1}\bar{\mathcal{S}}^{(1)}_{2k}
+ (1-\gamma_1) \bar{\mathcal{S}}^{(1)}_{2k} \mathcal{S}^{(1)}_{2k+1}\Big) \nonumber   \\
& + J\Big((1+\gamma_2)\mathcal{S}^{(2)}_{2k-1} \bar{\mathcal{S}}^{(2)}_{2k}
+ (1-\gamma_2)\bar{\mathcal{S}}^{(2)}_{2k} \mathcal{S}^{(2)}_{2k+1}\Big) \Big].
\label{Hamiltonian-a}
\end{align}
Further analysis of this Hamiltonian can be done following the same steps as for the $p$-$q$ model. Introducing coherent states parametrized by the matrix $\Omega$ leads to the following Lagrangian:
\begin{align}
{\cal L}_a &= \sum_{l,k} (-1)^k {\cal L}_B[\Omega_k^{(l)}] - \frac{J_\perp N^2}{4} \str \sum_k  \Omega_k^{(1)} \Omega_k^{(2)} \nonumber \\
&\quad + \frac{J N^2}{4} \str \sum_{l,k}\big[1 - (-1)^k \gamma_l \big] \Omega_k^{(l)} \Omega_{k+1}^{(l)}.
\label{Lagrangian-4(a)}
\end{align}
Notice that this Lagrangian differs from Eq. (\ref{Lagrangian}) by the sign of the interchain coupling, by the relative sign of the Berry phase terms of the two chains, and by the staggering pattern of the dimerization parameters $\gamma_l$. Formally, if we substitute $\Omega_k^{(l)} \to (-1)^l \Omega_k^{(l)}$, Eq. (\ref{Lagrangian-4(a)}) takes the form of Eq. (\ref{Lagrangian}) except that the dimerization parameters are replaced by
\begin{align}
\gamma_l &\to {\tilde \gamma}_l \equiv (-1)^{l-1} \gamma_l.
\end{align}
The corresponding substitutions
\begin{align}
Q_l &\to (-1)^l Q_l, & L_l &\to (-1)^l L_l, & \gamma_l &\to  {\tilde \gamma}_l
\end{align}
can be done  in the continuum action (\ref{coupNLSM}), and this leads to the action [recall the convention (\ref{shorthand})]
\begin{align}
S_a &= -\frac{N}{4} \int \!\! \sum_l L_l Q_l \partial_\tau Q_l \nonumber \\
& -\frac{J N^2 a}{4} \int \! \sum_l \Big[(1 + {\tilde \gamma}_l)\partial_x Q_l \big(\partial_x Q_l - 2 L_l \big) + 2 L_l^2\Big] \nonumber \\
&\quad -\frac{J_\perp N^2 a}{8} \int \!\! \Big[\frac{1}{a^2}\big(Q_1 + Q_2\big)^2 + \big(L_1 + L_2\big)^2 \Big].
\label{action-S_a}
\end{align}
In the limit $J_\perp = 0$ this action describes two decoupled superspin chains which are critical when both ${\tilde \gamma}_l = 0$. The interchain coupling is relevant, as before, and leads to the locking of the fields $Q_1 = - Q_2 = Q$ in the infrared. Once this happens, the $L$ fields can be integrated out, which gives again the sigma model action (\ref{NLSMAction}) with the couplings (\ref{conductivities-two-leg-ladder}). When the $\gamma_1 = \gamma_2 = \gamma$, we recover the conductivities (\ref{sigma-xy-staggered-two-legs}) and (for $N=1$) the critical lines (\ref{critical-lines-sigma model}). Thus, the phase diagram of the model (\ref{Hamiltonian-a}) should be qualitatively similar to the one of the $p$-$q$ model, with the critical point at $J_\perp/J=0$, $\gamma = 0$ giving rise to two critical lines for $J_\perp \neq 0$.

\begin{figure}[t]
\centerline{\includegraphics[width=80mm,angle=0,clip]{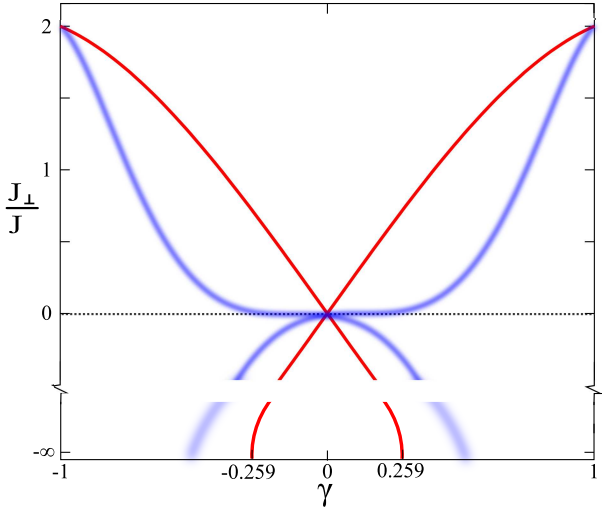}}
\caption{(Color online) Combined phase diagram of random networks and their spin-$\frac{1}{2}$ counterparts. Phase boundaries (critical states) of spin-$\frac{1}{2}$ models Eqs. (\ref{SSL}) and (\ref{spinHamA}), corresponding to $J_{\perp}/J>0$ and $J_{\perp}/J<0$, respectively, are plotted in red, whereas phase boundaries (extended states) of the $p$-$q$ model, $J_{\perp}/J>0$, and of the network Fig. \ref{diffnets}(a), $J_{\perp}/J<0$, are plotted in blue.} \label{combphasediag}
\end{figure}

This conclusion can be again substantiated by the analysis of the su(2) spin ladder that is analogous to the model (\ref{Hamiltonian-a}). The corresponding Hamiltonian is
\begin{align}
\label{spinHamA}
H_a &= J\sum_{l=1,2}\sum_k[1+(-1)^k\gamma] {\bf S}^{(l)}_k {\bf S}^{(l)}_{k+1}
-J_\perp\sum_k {\bf S}^{(1)}_k {\bf S}^{(2)}_k,
\end{align}
which differs from Eq. (\ref{SSL}) by the pattern of staggering and the sign of the interchain coupling, which is now ferromagnetic. This spin-$\frac{1}{2}$ model has been previously studied both analytically [\onlinecite{TS}] (bosonization and sigma model mapping) and numerically [\onlinecite{AMDS}] (DMRG). Here we briefly outline its basic properties in the limits of weak coupling, $J_\perp/J \ll 1$, and strong coupling, $J_\perp/J \gg 1$. These properties, combined with our analytical arguments for the weak coupling regime, will lead us to the unified quantum phase diagrams for the models (\ref{SSL}) and (\ref{spinHamA}), as well as the $p$-$q$ model (\ref{Hamiltonian-p-q}) and the model (\ref{Hamiltonian-a}), see Fig.~\ref{combphasediag}.

In the weak coupling regime $J_\perp/J \ll 1$, the system described by Eq. (\ref{spinHamA}) consists of two effectively decoupled $S=\frac{1}{2}$ dimerized Heisenberg chains. These chains are gapped unless $\gamma=0$. This is the same behavior as in the model (\ref{SSL}). Adapting the theory of Ref.~[\onlinecite{WN}] for the model (\ref{SSL}), it can be demonstrated that the two systems, (\ref{SSL}) and (\ref{spinHamA}), are {\it dual} to each other asymptotically exactly, in the limit $J_\perp/J\rightarrow 0$. This duality means that the analytical behavior of critical lines near the point $\gamma=0$ for two models coincides and is given by Eq. (\ref{critline2}) (with $J_\perp$ replaced by its absolute value $|J_\perp|$). This allows us to present the quantum phase diagrams of the two models in a single Fig. \ref{combphasediag}.

In the case of strong interchain coupling $J_\perp/J \gg 1$, on the other hand, the spins on each rung form a single spin-1, and the system (\ref{spinHamA}) turns into a single dimerized $S=1$ spin chain. Analytical [\onlinecite{Hal, AffHal, TNHS}] and numerical [\onlinecite{TNHS}, \onlinecite{KTH}] studies of the dimerized $S=1$ spin chain established that its spectrum is gapped for all values of $\gamma$ except for critical points at non-zero values $|\gamma|=\gamma_0=0.259$. A Haldane phase is realized in the region $|\gamma|<\gamma_0$, whereas the region $|\gamma|>\gamma_0$ corresponds to a dimer phase. Based on this basic knowledge, we sketch the phase diagram of Eq. (\ref{spinHamA}) in the lower panel of Fig. \ref{combphasediag}, corresponding to negative (ferromagnetic) interchain coupling. Our critical curve, $\gamma_c(J_\perp/J)$, starts from $\gamma_c(0)=0$ and at large negative arguments asymptotically approaches to $\gamma_c=\pm\gamma_0$. For more details on analytical shape of the critical lines in intermediate region $J_\perp/J\sim1$ the reader may consult Ref. [\onlinecite{AMDS}], where the model (\ref{spinHamA}) was studied numerically. We note in passing that the strong coupling limit of Eq. (\ref{SSL}) is not at all related to an $S=1$ spin chain because of the shifted dimerization patterns on constituting $S=\frac{1}{2}$ chains. This difference in strong coupling limit of Eq. (\ref{SSL}), where the spectrum is gapped for any $\gamma$, and the same limit of Eq. (\ref{spinHamA}), where there are two critical states no matter how strong the coupling gets, is clearly seen on the phase diagram Fig. \ref{combphasediag}.

%%%%%%%%%%%%%%%%%%%%%%%%%%%%%%%
%%%%%%%%%%%%%%%%%%

We believe that phase diagrams of the $p$-$q$ model and the network shown in Fig. \ref{diffnets}(a), as well as of their superspin counterparts (\ref{Hamiltonian-p-q}) and (\ref{Hamiltonian-a}), are qualitatively similar to the ones for the spin models (\ref{SSL}) and (\ref{spinHamA}), respectively. In particular, the shapes of the critical lines of the two networks (or, equivalently, of both superspin systems) in the vicinity of $\gamma=0$ are mirror images of each other relative to the horizontal axis. However, their analytical form is different from Eq. (\ref{critline2}). We reflect this fact by sketching the expected phase boundaries of the two networks in blue in Fig.~\ref{combphasediag}. For the $p$-$q$ model, neither our sigma model approach nor the earlier numerical simulations in Ref. [\onlinecite{MKR}] were able to resolve the actual shape of the critical line $\gamma_c(J_\perp/J)$ in the weak-mixing limit $J_\perp/J \ll 1$. As we saw above, this dependence should be captured by the coupled nonlinear sigma model (\ref{coupNLSM}) which is not solvable. For the numerical simulation, on the other hand, the difficulty resides in strong (quantum) fluctuations near the critical point $\gamma_c=0$, which are quite generic [\onlinecite{MKR}]. For the same reasons, the dependence $\gamma_c(J_\perp/J)$ in the weak-mixing limit was impossible to determine in models similar to the network in Fig. \ref{diffnets}(a), either from the coupled nonlinear sigma model arguments [\onlinecite{OGM}] or from numerics [\onlinecite{SorMac}]. Existing numerical results pertinent to the region of weak mixing, both for the $p$-$q$ model [\onlinecite{WeOrt}] and for the copropagating model of Ref. [\onlinecite{ChalOrt}], are performed in a classical limit where the models reduce to random walks. The resulting behavior of the critical line in the weak mixing regime for both models is $\gamma_c\propto(J_\perp/J)^{1/\beta}$ with $\beta$ close to $2$. According to this result and other qualitative arguments [\onlinecite{OGM}], we draw phase boundaries of the two networks [Figs. 4(a) and 4(d)], tangent to the horizontal axis at $\gamma=0$ in Fig. \ref{combphasediag}.

Let us now turn to the other two networks shown in Figs. \ref{diffnets}(b) and \ref{diffnets}(c). The corresponding superspin Hamiltonians are
\begin{align}
\mathcal{H}_b &= - \str\! \sum_k \Big[J_{\perp} \Big( \mathcal{S}^{(1)}_{2k-1} \mathcal{S}^{(2)}_{2k-1}
+ \bar{\mathcal{S}}^{(1)}_{2k} \bar{\mathcal{S}}^{(2)}_{2k}\Big) \nonumber \\
& + J\Big((1+\gamma_1) \mathcal{S}^{(1)}_{2k-1}\bar{\mathcal{S}}^{(1)}_{2k}
+ (1-\gamma_1) \bar{\mathcal{S}}^{(1)}_{2k} \mathcal{S}^{(1)}_{2k+1}\Big) \nonumber   \\
& + J\Big((1-\gamma_2)\mathcal{S}^{(2)}_{2k-1} \bar{\mathcal{S}}^{(2)}_{2k}
+ (1+\gamma_2)\bar{\mathcal{S}}^{(2)}_{2k} \mathcal{S}^{(2)}_{2k+1}\Big) \Big],
\label{Hamiltonian-b}
\\
\mathcal{H}_c &= - \str\! \sum_k \Big[J_{\perp} \Big( \mathcal{S}^{(1)}_{2k-1} \bar{\mathcal{S}}^{(2)}_{2k-1}
+ \bar{\mathcal{S}}^{(1)}_{2k} \mathcal{S}^{(2)}_{2k}\Big) \nonumber \\
& + J\Big((1+\gamma_1) \mathcal{S}^{(1)}_{2k-1}\bar{\mathcal{S}}^{(1)}_{2k}
+ (1-\gamma_1) \bar{\mathcal{S}}^{(1)}_{2k} \mathcal{S}^{(1)}_{2k+1}\Big) \nonumber   \\
& + J\Big((1+\gamma_2)\bar{\mathcal{S}}^{(2)}_{2k-1} \mathcal{S}^{(2)}_{2k}
+ (1-\gamma_2)\mathcal{S}^{(2)}_{2k} \bar{\mathcal{S}}^{(2)}_{2k+1}\Big) \Big].
\label{Hamiltonian-c}
\end{align}
The Hamiltonian (\ref{Hamiltonian-b}) differs from Eq. (\ref{Hamiltonian-a}) by the change in the sign of $\gamma_2$. The same sign difference distinguishes Eq. (\ref{Hamiltonian-c}) from the Hamiltonian (\ref{Hamiltonian}). This leads to the change $\gamma_\pm \to \gamma_\mp$, which can be done directly in the appropriate actions in the continuum. For both models, this leads to the nonlinear sigma model action (\ref{NLSMAction}) with conductivities
\begin{align}
\sigma_{xx} &= \frac{N}{\mu} \sqrt{\mu - \mu \gamma_+^2 - \gamma_-^2}, & \sigma_{xy} &=  -\frac{N\gamma_-}{\mu}.
\end{align}
An important consequence is that when $\gamma_1 = \gamma_2 = \gamma$, the $\theta$ angle vanishes, which makes the models massive for any nonzero values of $J_\perp$. For the network models in Figs. \ref{diffnets}(b) and \ref{diffnets}(c) this implies the absence of extended states for any values of the parameters (except in the limit of totally decoupled critical CC networks).

This conclusion is again corroborated by the similarity with the su(2) spin-$\frac{1}{2}$ ladders with the Hamiltonians
\begin{align}
H_b &= J\sum_{l,k} [1+(-1)^{k+l}\gamma] {\bf S}^{(l)}_k {\bf S}^{(l)}_{k+1} - J_\perp \sum_k {\bf S}^{(1)}_k {\bf S}^{(2)}_k,
\label{spinHamB}
\\
H_c &= J\sum_{l,k} [1+(-1)^k\gamma] {\bf S}^{(l)}_k{\bf S}^{(l)}_{k+1} + J_\perp \sum_k {\bf S}^{(1)}_k {\bf S}^{(2)}_k,
\label{spinHamC}
\end{align}
corresponding to the networks in Figs. \ref{diffnets}(b) and \ref{diffnets}(c), respectively. Indeed, both models (\ref{spinHamB}) and (\ref{spinHamC}) exhibit gapped spectra for all (positive) values of couplings, that is, they do not undergo any quantum phase transitions. This follows from the results of previous studies [\onlinecite{KimFurLee}, \onlinecite{TNHS}, \onlinecite{Chitov}]. We have drawn the same conclusion from an analysis based on mapping of Eqs. (\ref{spinHamB}) and (\ref{spinHamC}) onto $O(3)$ nonlinear sigma models.

\subsection{Physical implications}

In contrast to the counterpropagating $p$-$q$ model in Figs. \ref{fig:Network} and \ref{diffnets}(d), the copropagating random network shown in Fig. \ref{diffnets}(a) does not capture the peculiarities  of electron motion in vanishing magnetic field. Instead, the latter describes the strong-field integer quantum Hall effect in a double-layer system [\onlinecite{SorMac}, \onlinecite{GramRaikh}], if one identifies the two CC components with the two layers. Hence, our phase diagram Fig. \ref{combphasediag} represents the repulsion of energies of delocalized states predicted for such systems. In spite of difficulties in numerical simulations of Ref. [\onlinecite{SorMac}] in the weak-mixing regime $|J_{\perp}/J| \ll 1$, their phase diagram is in overall good agreement with ours.

The model of Fig. \ref{diffnets}(a) is also the random network description of the theory of splitting of delocalized states due to the valley mixing in graphene, put forward in Ref. [\onlinecite{OGM}]. The two channels copropagating on each link of Fig. \ref{diffnets}(a) describe the quasiclassical drift of guiding centers of Dirac particles belonging to inequivalent valleys, within a single Landau level of the disordered graphene in a strong magnetic field. Scattering between these channels corresponds to the mixing of states between different valleys, with an amplitude $p \propto J_{\perp}$. According to our results and in agreement with Ref. [\onlinecite{OGM}], the valley mixing splits the single delocalized state into two delocalized states with energy difference $\Delta E$, and with an intermediate plateau of Hall couductivity in between. According to the phase diagram shown in Fig. \ref{combphasediag}, for weak valley mixing $J_{\perp}/J \ll 1$, the energy splitting $\Delta E \propto J \gamma_c(J_{\perp}/J)$, is strongly sensitive to the mixing strength.

In Ref. [\onlinecite{LeeChalker}], a two-channel random network model was introduced to describe electron states in spin-degenerate Landau levels. In that network, the two channels correspond to the two possible spin orientations, and scattering between these channels is included by replacing the link phases with U(2) scattering matrices. On each link, these matrices are randomly and independently chosen uniformly over the Haar measure. Geometrically, the network model for spin-degenerate Landau levels is very similar to the network in Fig.~\ref{diffnets}(a). The only difference between the two networks is in the mixing strength $p$ of the fluxes on the links. This strength is random in the network of Ref. [\onlinecite{LeeChalker}], with the distribution function $Q(p)=2/(\pi\sqrt{1-p^2})$. The mean value of $p$ is $\langle p\rangle=2/\pi$, which corresponds to strong mixing.

The authors of Ref. [\onlinecite{LeeChalker}] numerically determined the phase diagram of their two-channel network and found that it had two critical lines supporting delocalized states. This is consistent with our phase diagram in Fig. \ref{PhaseDiag} (the negative $J_\perp$ region). The correspondence may be understood as follows. We can modify the distribution of the U(2) scattering matrices to make it anisotropic in such a way that all mixing parameters $p$ are the same along a vertical zig-zag (independent of the imaginary time), but randomly distributed in the spatial direction with the distribution $Q(p)$. Such ``quenched'' disorder can be treated within our supersymmetric approach, and leads to the superspin ladder with the Hamiltonian (\ref{Hamiltonian-a}) with random positive $J_{\perp}$. In this case, due to the large value of $\langle p\rangle$, we will have $\langle J_{\perp}\rangle/J \gtrsim 1$. We can argue that the randomness in $J_\perp$, as long as it preserves the ferromagnetic nature of the coupling, is irrelevant, and does not modify the qualitative features of the phase diagram.

Reference [\onlinecite{LeeChalker}] introduced another copropagating network to describe the localization in random magnetic field. This network is geometrically the same as our Fig. \ref{diffnets}(b), but with randomness in the mixing strength described by the distribution $Q(p)$ [since, again, the mixing was modeled by U(2) scattering matrices uniformly distributed over the Haar measure]. The authors found that model to have no extended states, similarly to the network of Fig. \ref{diffnets}(b). We can understand this by considering the anisotropic modification of the random magnetic field network described above. This leads to the superspin ladder (\ref{Hamiltonian-b}) but with random values of the couplings $J_\perp$. We do not expect such randomness to change the nature of the phase diagram, so that all the states in the corresponding network model should be localized, in full agreement with the results of Ref. [\onlinecite{LeeChalker}].

\section{Conclusions}
\label{sec:conclusions}

We have considered a minimal two-channel network model (the so-called $p$-$q$ model) for the integer quantum Hall effect in weak magnetic fields and the phenomenon of ``levitation'' of extended states within Landau levels. Using the supersymmetry method, we have mapped this network to a staggered superspin ladder. Under this mapping, the rung-dimer phase of the ladder corresponds to the Anderson insulator, while the leg-singlet phases correspond to the quantum Hall insulator phases of the low-field quantum Hall phase diagram.

We have analyzed the shape of the critical lines separating the localized phases using a combination of the effective field theory for the superspin ladder, the nonlinear sigma model with a topological term, and the intuition gained from the study of analogous ladder with $S=\frac{1}{2}$ su(2) spins. This analysis also demonstrated that the transitions between localized phases of the $p$-$q$ model are in the same universality class as the usual integer quantum Hall transitions modeled by the CC network model.

Using the supersymmetry method, we have also considered other two-channel network models. These models were previously proposed to describe quantum Hall effects in spin-degenerate Landau levels, in graphene with mixing between valleys, and localization of electrons in a random magnetic field. Our results are in complete agreement with previous findings about the localization behavior of these models.

\section{Acknowledgements}

We are grateful to M. Raikh and A. Mirlin for useful discussions. This work was supported by the BSF through Grant No. 2010030 and by the NSF through Grant No. DMR-1105509. The research of V.M. was supported by the DOE Office of Basic Energy Sciences, Grant No. DE-AC02-07CH11358, and by the Research Corporation for Science Advancement.

\appendix

\section{SUSY formalism}
\label{sec:appendix-SUSY}

In this appendix we provide details of the SUSY formalism for the $p$-$q$ network model. We start with algebraic preliminaries about certain irreps of the Lie superalgebra u$(1,1|2)$. We then construct the second-quantized supersymmetric form of the transfer matrices, average them over the disorder, and consider the time-continuum limit.

\subsection{Irreps of u$(1,1|2)$}
\label{sec:appendix-SUSY-1}

We introduce four pairs of creation and annihilation operators $c_m$, $c_m^*$ on an up half-channel, with a composite index $m = (s, S)$, where $s = R,A$ corresponds to the retarded/advanced sector, and $S = B, F$ corresponds to the boson/fermion sector. Explicitly, these operators are given by
\begin{align}
c_{R,B} &= b_R, & c_{R,F} &= f_R, & c_{A,B} &= -b_A^\dagger, & c_{A,F} &= f_A^\dagger, \nonumber \\
c_{R,B}^* &= b_R^\dagger, & c_{R,F}^* &= f_R^\dagger, & c_{A,B}^* &= b_A, & c_{A,F}^* &= f_A.
\label{upfb}
\end{align}
Here, $b$ and $f$ are the canonical bosonic and fermionic annihilation operators, satisfying the usual (anti)commutation relations. We define the parity $|m|$ as $0$ for bosonic operators and $1$ for fermionic operators. Then, the (anti)commutation relations acquire the form
\begin{align}
\label{CommutationRel}
\llbracket c^{\vphantom *}_m, c^{*}_n \rrbracket \equiv c^{\vphantom *}_m c^{*}_n - (-1)^{|m||n|} c^{*}_n c^{\vphantom *}_m = \delta_{mn}.
\end{align}

The 16 generators of u$(1,1|2)$ appear as bilinears in bosons and fermions:
\begin{align}
\mathcal{S}_{mn} = c^{\vphantom *}_m c^{*}_n - \frac{1}{2}\delta_{mn}.
\label{higweirep}
\end{align}
It is convenient to arrange the generators in a $4 \times 4$ supermatrix $\mathcal{S}$, and refer to them collectively as components of a ``superspin.'' For such supermatrices we define the supertrace ``str'' as
\begin{align}
\str A = \sum_m (-1)^{|m|} A_{mm}.
\end{align}
With this notation we have
\begin{eqnarray}
\str \mathcal{S} &=& n_{b_R} + n_{f_R} - n_{b_A} - n_{f_A},
\label{strS}
\end{eqnarray}
where the number operators are given by the usual expressions
\begin{align}
n_{b_s} &= b_s^\dagger b_s^{\vphantom \dagger}, & n_{f_s} &= f_s^\dagger f_s^{\vphantom \dagger}, & s &= R, A.
\end{align}

The components of the superspin $\cal S$ act in the Fock space associated with the operators $b_s$ and $f_s$.  Applying these generators to the vacuum $\ket{0}$ defined as usual by
\begin{align}
b_R\ket{0} = b_A\ket{0} = f_R\ket{0} = f_A\ket{0} =0,
\label{defvacuum}
\end{align}
we obtain an irreducible representation (irrep) of the Lie superalgebra u$(1,1|2)$ which we denote by $R$. This irrep can also be obtained by imposing the constraint $\str \mathcal{S} = 0$ in the Fock space. $R$ is a highest-weight representation with the vacuum $\ket{0}$ playing the role of the highest-weight vector. The components of the highest weight $\lambda$ are given by the eigenvalues of the Cartan generators [diagonal components of the superspin (\ref{higweirep})] in the vacuum state:
\begin{align}
\lambda =\frac{1}{2} \left(1,1,-1,-1 \right).
\label{highweight}
\end{align}

On the down half-channels, we introduce another set of bosonic and fermionic operators:
\begin{align}\label{downfb}
\bar{c}_{R,B} &= \bar{b}_R^\dagger, & \bar{c}_{R,F} &= \bar{f}_R^\dagger, & \bar{c}_{A,B} &= -\bar{b}_A,
& \bar{c}_{A,F} &= \bar{f}_A, \nonumber \\
\bar{c}_{R,B}^* &= \bar{b}_R, & \bar{c}_{R,F}^* &= \bar{f}_R, & \bar{c}_{A,B}^* &= \bar{b}_A^\dagger, & \bar{c}_{A,F}^* &= \bar{f}_A^\dagger.
\end{align}
Internal consistency of the SUSY description (cancellation of contributions from vacuum loops) requires the states with odd numbers of ``up'' fermions to have ``negative norms,'' which is achieved by assigning the following noncanonical (anti)commutation relations:
\begin{align}
\label{CommutationRel-ConjRep}
\llbracket \bar{c}^{\vphantom *}_m,\bar{c}^{*}_n \rrbracket \equiv
\bar{c}^{\vphantom *}_m \bar{c}^{*}_n - (-1)^{|m||n|} \bar{c}^{*}_n \bar{c}^{\vphantom *}_m
= -\delta_{mn}.
\end{align}
Then, the number operators for the down particles are
\begin{align}
n_{\bar{b}_s} &= \bar{b}_s^\dagger \bar{b}_s^{\vphantom \dagger}, & n_{\bar{f}_s} &= -\bar{f}_s^\dagger \bar{f}_s^{\vphantom \dagger}, & s &= R, A.
\end{align}

The operators $\bar{c}$, $\bar{c}^*$ provide another realization of the generators of u$(1,1|2)$:
\begin{align}
\bar{\mathcal S}_{mn} = - \bar{c}^{\vphantom *}_m \bar{c}^{*}_n - \frac{1}{2}\delta_{mn}.
\end{align}
When acting on the vacuum $\ket{\bar{0}}$ (annihilated by $\bar{b}_s$ and $\bar{f}_s$), these generators form another representation $\bar R$ of u$(1,1|2)$.  This is the lowest-weight representation dual to $R$. The lowest weight $\bar \lambda$ is opposite to the highest-weight $\lambda$ (as it should be for dual representations):
\begin{align}
{\bar \lambda} = - \lambda  = - \frac{1}{2} \left(1,1,-1,-1 \right).
\label{lowweight}
\end{align}

%%%```````````````````````````````````
\begin{figure}[t!]
\centering
\includegraphics[width=0.8\columnwidth]{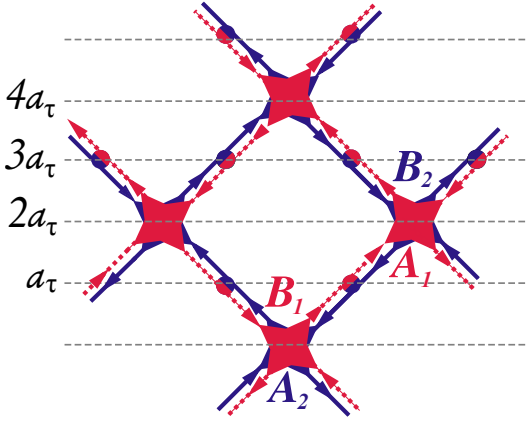}
\caption{(Color online) The $p$-$q$ model.  The horizontal dashed lines denote time slices at the values of time coordinate  $\tau$ which are integer multiples of $a_\tau$.}
\label{fig:Network-Time}
\end{figure}
%%%```````````````````````````````````

\subsection{Transfer matrices in SUSY form}
\label{sec:appendix-SUSY-2}

In our network model, each elementary $2 \times 2$ scattering matrix
\begin{align}
S(t_X) =
\begin{pmatrix} \alpha_X & \beta_X \\ \gamma_X & \delta_X \end{pmatrix} \equiv
\begin{pmatrix} \sqrt{1-t_X^2} & t_X \\ - t_X & \sqrt{1-t_X^2} \end{pmatrix},
\end{align}
where $X$ can take values $P$, $A_1$, $A_2$, $B_1$, and $B_2$, describes scattering between up- and down-going fluxes. Using bosonic and fermionic operators (\ref{upfb}) and (\ref{downfb}), the second-quantized form of the transfer matrix associated with $S_X$ is [\onlinecite{GruzReadSach}, \onlinecite{GruzLudRead}]
\begin{align}
V_X = \prod_{s = R, A} & \exp \big[\beta_X \big(b_s^\dagger \bar{b}_s^\dagger + f_s^\dagger \bar{f}_s^\dagger\big) \big] \alpha_X^{n_{b_s} + n_{f_s}} \delta_X^{n_{\bar{b}_s} + n_{\bar{f}_s}} \nonumber \\
\times & \exp \big[\gamma_X \big(\bar{b}_s b_s + \bar{f}_s f_s \big) \big].
\end{align}

We multiply all such transfer matrices to construct the total evolution operator of the network $U$ in the discrete time $\tau$. To be specific, let us consider a network which has the vertical size
\begin{align}
L_\tau = 4 N_\tau a_\tau,
\end{align}
where $N_\tau$ is the number of links in every vertical column, and $a_{\tau}$ denotes the elementary time interval, equal to the vertical separation between the middle points of two adjacent half-links.
%This is also one half of the vertical separation between the nearest rows of nodes, see Fig.
%\ref{fig:Network-Time}.
We choose the origin for the vertical $\tau$ axis in such a way that the $\tau$ coordinates of the middles of the links, where the scattering matrices $S(t_P)$ are located, are odd multiples of $a_\tau$. The $\tau$ coordinates of the $A$ and $B$ nodes are then even multiples of $a_\tau$ (see Fig. \ref{fig:Network-Time}). Different boundary conditions may be imposed in the $\tau$ direction. To be specific, and for simplicity, here we assume periodic boundary conditions.

In the horizontal direction we label all columns of links (and the corresponding bosons and fermions) by $i = 1, 2, \ldots, 2 N_x$. Upon taking the time-continuum limit, the columns of links will become sites of a superspin ladder, and the index $i$ will label the sites along the ladder. We also have to label fermions and bosons by the index $l = 1,2$ distinguishing the two subnetworks. The ordering of indices will be as follows: $b_{l, i, s}$, etc.

Using these conventions, a generic transfer matrix located at the discrete time $\tau$ and connecting bosons and fermions on sites $i$ and $i'$ can be written as
\begin{align}
V_X^{(l,l')}[\tau; i, i'] &= \!\! \prod_{s = R, A} \!\! \exp \big[\beta_X \big(b_{l, i, s}^\dagger
\bar{b}_{l', i', s}^\dagger + f_{l, i, s}^\dagger \bar{f}_{l', i', s}^\dagger\big) \big] \nonumber \\
& \quad \times \alpha_X^{n_{b_{l, i, s}} + n_{f_{l, i, s}}} \delta_X^{n_{\bar{b}_{l', i', s}} + n_{\bar{f}_{l', i', s}}} \nonumber \\
& \quad \times \exp \big[\gamma_X \big(\bar{b}_{l', i', s} b_{l, i, s} + \bar{f}_{l', i', s} f_{l, i, s} \big) \big].
\label{V_X}
\end{align}
Notice that the indices $l, i$ label up-going fluxes and the corresponding bosons and fermions. Since fermionic operators do not commute, the order of indices is important.

Next, for each $\tau$ we take the product of the transfer matrices with this time argument. We have the following products:
\begin{align}
U_1(\tau) &\equiv  \prod_{k=1}^{N_x} V_P^{(1,2)}[\tau; 2k-1,2k-1] V_P^{(2,1)}[\tau; 2k,2k], \nonumber\\
U_2(\tau) &\equiv \prod_{k=1}^{N_x} V_{A_1}^{(1,1)}[\tau; 2k-1,2k] V_{B_2}^{(2,2)}[\tau; 2k,2k-1], \nonumber\\
U_3(\tau) &\equiv \prod_{k=1}^{N_x - 1} V_{B_1}^{(1,1)}[\tau; 2k+1,2k] V_{A_2}^{(2,2)}[\tau; 2k,2k+1].
\end{align}
Then, for each $\tau$ that is a multiple of $4a_\tau$ we construct the evolution operator $U(\tau)$ for one time step $\Delta \tau = 4a_\tau$ by multiplying the above operators as
\begin{align}
U(\tau) &= U_3(\tau + 4a_\tau) U_1(\tau + 3a_\tau) U_2(\tau + 2a_\tau) U_1(\tau + a_\tau).
\end{align}
Finally, the full time evolution operator~$U$ is then given by the product
\begin{align}
U = {\text T}_\tau \prod_{n=0}^{N_{\tau}-1} U(4n a_\tau),
\end{align}
which is ordered with the earliest times at the right. The time ordering is indicated by ${\text T}_\tau$.

\subsection{Disorder average and time-continuum limit}
\label{sec:appendix-SUSY-3}

So far we have not included the random phases on the half-links. As is explained in Refs. [\onlinecite{GruzReadSach}, \onlinecite{GruzLudRead}], averaging over the random phases leads to the projection of the transfer matrices $V_X$ (and the evolution operator $U$) to the tensor product
\begin{align}
\bigotimes_{i = 1}^{2 N_x} (R_i \otimes \bar{R}_i)
\label{space-of-ladder}
\end{align}
of irreducible representations $R$ and $\bar{R}$, one of each for each horizontal coordinate $i$. This tensor product is the space of the superspin ladder, and we denote the projection onto it by $P$.

In the anisotropic time-continuum limit, all scattering parameters $t_X \ll 1$, and we can expand each transfer matrix (\ref{V_X}) in $t_X$. Projecting the result to the space (\ref{space-of-ladder}), one can show that
\begin{align}
\label{projectiontoSuperspins}
P V_X^{(l,l')}[\tau; i, i'] P &= 1 + t_X^2 \str \mathcal{S}^{(l)}_i \bar{\mathcal{S}}^{(l')}_{i'} + O(t_X^3),
\end{align}
where we have moved the indices distinguishing the sublattices to superscripts for convenience.  Multiplying such projected expansions for all transfer matrices entering the evolution operator $U$, we get
\begin{align}
[U] \approx 1 &+ \str \sum_n \Big[\sum_{k=1}^{N_x-1} \big(
t_{B_1}^2 \mathcal{S}^{(1)}_{2k+1} \bar{\mathcal{S}}^{(1)}_{2k} +
t_{A_2}^2 \mathcal{S}^{(2)}_{2k} \bar{\mathcal{S}}^{(2)}_{2k+1} \big) \nonumber \\
& + \sum_{k=1}^{N_x} \big(t_{A_1}^2 \mathcal{S}^{(1)}_{2k-1} \bar{\mathcal{S}}^{(1)}_{2k}
+ t_{B_2}^2 \mathcal{S}^{(2)}_{2k} \bar{\mathcal{S}}^{(2)}_{2k-1} \big)
\nonumber \\
& + \sum_{k=1}^{N_x} 2 t_P^2 \big(\mathcal{S}^{(1)}_{2k-1}
\bar{\mathcal{S}}^{(2)}_{2k-1} + \mathcal{S}^{(2)}_{2k}
\bar{\mathcal{S}}^{(1)}_{2k}\big)\Big].
\end{align}
Let us now introduce the following notation for the coupling constants $J_X = t_X^2$ for $X = A_l, B_l$ and $J_\perp = 2t_P^2$. Next, we replace the sum over the time slices $n$ by the integral over $\tau$, and also rescale the ``imaginary time'' as
\begin{align}
\tau &\to \frac{\tau}{4 a_\tau}, & L_\tau \to \beta = \frac{L_\tau}{4 a_\tau} = N_\tau.
\end{align}
This results in the disorder averaged ``evolution operator''
\begin{align}
[U] &\approx \exp{\Big( -\int_0^\beta \!\! d\tau\, \mathcal{H}_{1D} \Big)}, \\
\mathcal{H}_{1D} &= -\str \sum_k \Big[J_{A_1} \mathcal{S}^{(1)}_{2k-1} \bar{\mathcal{S}}^{(1)}_{2k}
+   J_{B_1} \bar{\mathcal{S}}^{(1)}_{2k} \mathcal{S}^{(1)}_{2k+1} \nonumber \\
& \qquad \quad \quad + J_{B_2} \bar{\mathcal{S}}^{(2)}_{2k-1} \mathcal{S}^{(2)}_{2k}
+ J_{A_2} \mathcal{S}^{(2)}_{2k} \bar{\mathcal{S}}^{(2)}_{2k+1} \nonumber \\
& \qquad \quad \quad + J_\perp \big(\mathcal{S}^{(1)}_{2k-1} \bar{\mathcal{S}}^{(2)}_{2k-1}
+ \bar{\mathcal{S}}^{(1)}_{2k} \mathcal{S}^{(2)}_{2k} \big)\Big].
\label{eq:1D-Hamiltonian}
\end{align}
%\end{widetext}

We now focus on the specific choice of couplings (\ref{eq:anisotropic-p-q}), in which case we have $(i = 1, 2)$
\begin{align}
& t_{A_l}^2 + t_{B_l}^2  = \epsilon,  &&  t_{A_l}^2 - t_{B_l}^2 = \epsilon \gamma_l, \nonumber \\
& t_{A_l}^2 = \frac{1+\gamma_l}{2} \epsilon, && t_{B_l}^2 = \frac{1-\gamma_l}{2} \epsilon.
\end{align}
Then denoting $J = \epsilon/2$ we have
\begin{align}
J_{A_l} &= J(1+\gamma_l), & J_{B_l} &= J(1-\gamma_l),
\end{align}
and the one-dimensional Hamiltonian (\ref{eq:1D-Hamiltonian}) reduces to the form given in Eq. (\ref{Hamiltonian}).

\section{Coherent states and path integral for u$(n,n|2n)$}
\label{sec:appendix-coherent-states}

In this appendix we construct the coherent states (see Ref. [\onlinecite{Perelomov}] for a general discussion) for the group U$(n,n|2n$) following Refs. [\onlinecite{Wiegmann}, \onlinecite{Read+Sachdev}] (see also Refs. [\onlinecite{Zirnbauer}, \onlinecite{BalentsFisherZirnbauer}]) with some minor differences. We then use the constructed coherent states to represent the partition function of a superspin system as a functional integral.

\subsection{Highest-weight irreps of u$(n,n|2n)$ and U$(n,n|2n)$}

Let us start by introducing some notation. We generalize Eq. (\ref{upfb}) by introducing $n$ copies of fermions and bosons, labeling them by the replica index, $\alpha=1,\ldots, n$. We also add $N$ channels in each direction along each link, labeled by the channel index, $i=1,\ldots, N$. The number of channels $N$ will ultimately be seen as the large parameter that controls the sigma model derivation, and will be related to the bare value of the longitudinal conductivity.

We denote the bosonic and fermionic operators for an up-going half-link as $c_{iI}$, where we combine the retarded/advanced index $s$, the replica index $\alpha$, and the fermion/boson index $S$ into the composite index $I$. These are the same operators as in Eq. (\ref{upfb}) but with the addition of $i$ and $\alpha$ indices everywhere, and they satisfy the commutation relations
\begin{align}
\llbracket c_{iI}^{\vphantom *}, c_{jJ}^* \rrbracket \equiv  c_{iI}^{\vphantom *} c_{jJ}^* -
(-1)^{|I||J|} c_{jJ}^* c_{iI}^{\vphantom *} = \delta_{ij}\delta_{IJ}.
\label{CommRels}
\end{align}
As before, $|I|=0$ for a boson and $|I|=1$ for a fermion.

The commutation relations (\ref{CommRels}) are preserved under canonical transformations that form the group GL$(2n|2n)$. Elements $g$ of GL$(2n|2n)$ are $4n \times 4n$ complex supermatrices acting on (the second indices of) the operators $c$ and $c^*$ as
\begin{align}
c_{iI} &\to (g^{-1})_{IJ}^{\vphantom *} c_{iJ}^{\vphantom *}, & c_{iI}^* &\to c_{iJ}^* g_{JI}^{\vphantom *}.
\label{g-action}
\end{align}
The group $G \equiv \text{U}(n,n|2n)$, a real form of GL$(2n|2n)$, consists of matrices satisfying the ``reality'' condition:
\begin{align}
g^{-1} &= \eta g^\dagger \eta, & \eta &= \diag(I_n, I_n, -I_n, I_n).
\label{reality-group}
\end{align}
The matrix $\eta$ determines a bilinear form (in the graded space ${\mathbb C}^{2n,2n}$) preserved by $G$. (Note that we have adopted the so-called ``retarded-advanced'' notation, in which the first $2n$ components of a supervector, both bosonic and fermionic, are retarded.) If we parametrize $g = \exp (i {\cal A})$, where $i {\cal A} \in \text{u}(n, n, |2n)$, then the reality condition translates to ${\cal A}^\dagger = \eta {\cal A} \eta$. Furthermore, if we decompose ${\cal A}$ into $2n \times 2n$ blocks as
\begin{align}
{\cal A} = \begin{pmatrix} A_{RR} & A_{RA} \\ A_{AR} & A_{AA} \end{pmatrix},
\end{align}
then the blocks satisfy
\begin{align}
A_{RR}^\dagger &= A_{RR}, & A_{RA}^\dagger &= - \Sigma_z A_{AR}, & A_{AA}^\dagger &= \Sigma_z A_{AA} \Sigma_z,
\label{reality-blocks}
\end{align}
where $\Sigma_z = \diag (I_n, -I_n)$ is a diagonal $2n \times 2n$ matrix.

The $16n^2$ generators of u$(n,n|2n)$, the Lie superalgebra of $G$, can be constructed as the following bilinears in bosons and fermions:
\begin{align}
\mathcal{S}_{IJ} = \sum_{i=1}^N c_{iI}^{\vphantom *} c_{iJ}^* - \frac{N}{2}\delta_{IJ}.
\end{align}
As before, it is convenient to arrange these generators in the form of a $4n \times 4n$ matrix $\mathcal S$ and refer to this matrix as a superspin. The superspin components generate a highest-weight representation $R_N$ when acting on the vacuum $\ket{0}$ of all the bosons and fermions. The diagonal blocks ${\cal S}_{ss}$ (in the retarded/advanced ordering) of the superspin act on the vacuum as
\begin{align}
({\mathcal S}_{RR})_{\alpha,S;\alpha',S'}\ket{0} &= \frac{N}{2} \delta_{\alpha,\alpha'} \delta_{S,S'} \ket{0}, \nonumber \\
({\mathcal S}_{AA})_{\alpha,S;\alpha',S'}\ket{0} &= -\frac{N}{2} \delta_{\alpha,\alpha'} \delta_{S,S'} \ket{0}.
\label{block-action}
\end{align}
This implies, in particular, that the vacuum is the highest-weight state with the weight
\begin{align}
\lambda_N = \frac{N}{2}
(\underbrace{1,\ldots,1}_{\mbox{\scriptsize $2n$ times}}, \underbrace{-1,\ldots,-1}_{\mbox{\scriptsize $2n$ times}}).
\label{high}
\end{align}
The expectation value of $\mathcal{S}$ in the vacuum is
\begin{align}
\bra{0} \mathcal{S} \ket{0} = \frac{N}{2} \Lambda,
\label{vacuum-superspin}
\end{align}
where $\Lambda$ is the diagonal matrix
\begin{align}
\Lambda = \diag (I_{2n}, -I_{2n}).
\label{Lambda}
\end{align}

We can construct a unitary representation of $G$ which corresponds to $R_N$. To this end, for every element $g \in G$ we consider the operator
\begin{align}
T_g &= \exp \left[ \str(\ln g)\mathcal{S} \right] = \exp \big[ c_{iI}^* (\ln g)_{IJ}^{\vphantom *} c_{iJ}^{\vphantom *} \big] (\sdet g)^{N/2}.
\label{T_g}
\end{align}
These operators form a representation of $G$ in the Fock space by virtue of the following relations:
\begin{align}
T_g^{\vphantom 1} c_{iI} T_g^{-1} &= (g^{-1})_{IJ}^{\vphantom *} c_{iJ}^{\vphantom *}, &
T_g^{\vphantom 1} c_{iI}^* T_g^{-1} &= c_{iJ}^* g_{JI}^{\vphantom *}.
\label{Tg-adjoint-action}
\end{align}
This representation in terms of $T_g$ is unitary:
\begin{align}
T_g^\dagger &= T_{g^{-1}} = T_g^{-1}.
\end{align}

\subsection{Coherent states}

Now, we are ready to introduce the coherent states. According to the general scheme of Ref. [\onlinecite{Perelomov}], we should identify the subgroup $H$ of $G$ which stabilizes the vacuum $\ket{0}$. If a general element of $G$ is decomposed as
\begin{align}
g = \exp i \begin{pmatrix} A_{RR} & A_{RA} \\ A_{AR} & A_{AA} \end{pmatrix},
\end{align}
then the required subgroup $H$ is the group $\text{U}(n|n) \times \text{U}(n|n)$ of elements of the form
\begin{align}
h = e^{i \, \diag (A_{RR}, A_{AA})}.
\end{align}
Indeed, according to the general formula (\ref{T_g})
\begin{align}
T_h \ket{0} &= \exp\big[i \, \str(A_{RR} {\mathcal S}_{RR} + A_{AA} {\mathcal S}_{AA}) \big] \ket{0} \nonumber \\
&= e^{i \frac{N}{2} \str (A_{RR} - A_{AA})} \ket{0},
\end{align}
where we used the action (\ref{block-action}) of the diagonal blocks of the superspin on the vacuum.

The coherent states are defined as
\begin{align}
\ket{\omega} = T_{g_\omega} \ket{0},
\end{align}
where a convenient choice of $g_\omega$ is
\begin{align}
g_\omega = e^{i {\cal A}_\omega} = \exp i \begin{pmatrix} 0 & \omega \\ -\Sigma_z \omega & 0 \end{pmatrix},
\end{align}
where $\omega \in \text{GL}(n|n)$ is an arbitrary $2n \times 2n$ supermatrix. Thus, we have used only the generators of u$(n,n|2n)$ which are not generators of $\text{u}(n|n) + \text{u}(n|n)$, and remembered the reality condition (\ref{reality-blocks}). The coherent states $\ket{\omega}$ have several important properties. First of all, the expectation value of the superspin in a coherent state is
\begin{align}
\bra{\omega} {\cal S}_{IJ} \ket{\omega} &= \frac{N}{2} (g_\omega)_{IK} \Lambda_{KL} (g_\omega^{-1})_{LJ},
\end{align}
where we used (\ref{vacuum-superspin}) and (\ref{Tg-adjoint-action}). Thus,
\begin{align}
\bra{\omega} {\cal S} \ket{\omega} &= \frac{N}{2} \Omega, & \Omega
&= g_\omega^{\vphantom 1} \Lambda g_\omega^{-1}. \label{Omega}
\end{align}

The matrix $\Omega$ obviously satisfies
\begin{align}
\Omega^2 = I_{4n}
\label{Omega-square}
\end{align}
and belongs to the supermanifold $G/H$. Denoting the $G$-invariant measure on this manifold by $d\Omega$, we also have the following resolution of identity in the representation space of $R_N$ expressing the completeness of the system of coherent states:
\begin{align}
P = \int \! d\Omega \, \ket{\omega} \bra{\omega}.
\label{projector-P}
\end{align}
We call this operator $P$ since in the total Fock space it plays the role of the projection operator to the space $R_N$. This is a local (a single superspin) version of the projection operator that appeared in Appendix \ref{sec:appendix-SUSY-3}.

\subsection{Functional integral}

Now, we have all the ingredients necessary to express the partition function for a single superspin as a functional integral. Breaking the imaginary-time interval $[0, \beta]$ into infinitesimal intervals, and inserting the resolution of identity (\ref{projector-P}) in the usual way, we get
\begin{align}
Z = \int\! {\cal D} \, \Omega \, e^{-S},
\end{align}
where the action is
\begin{align}
S &= \int_0^{\beta} \!\! d\tau \Big(\bra{\omega(\tau)} \partial_\tau \ket{\omega(\tau)} + {\cal H}[N \Omega(\tau)/2] \Big).
\label{action-1}
\end{align}
The matrices $\omega(\tau)$ and
\begin{align}
\Omega(\tau) = g_{\omega(\tau)}^{\vphantom 1} \Lambda g_{\omega(\tau)}^{-1}
\label{Omega-tau}
\end{align}
are periodic in $\tau$: $\omega(\beta) = \omega(0)$, $\Omega(\beta) = \Omega(0)$, and so trace closed trajectories in the corresponding spaces.

By standard manipulations the first term in Eq. (\ref{action-1}), the Berry phase term $S_B$, can be rewritten as
\begin{align}
S_B &= \frac{N}{4} \int_0^{\beta} \!\! d\tau \int_0^1 \!\! du \,
\str [\Omega(\tau, u) \partial_u \Omega(\tau, u) \partial_\tau \Omega(\tau, u)],
\label{Berry-phase}
\end{align}
where $\Omega(\tau, u)$ is a smooth homotopy between $\Omega(\tau, 0) = \Lambda$ and $\Omega(\tau, 1) = \Omega(\tau)$ defined similarly to (\ref{Omega-tau}) as
\begin{align}
\Omega(\tau, u) = g(\tau,u) \Lambda g^{-1}(\tau,u),
\label{Omega-tau-u}
\end{align}
where
\begin{align}
g(\tau, u) = e^{i u {\cal A}_{\omega(\tau)}} = \exp i u \begin{pmatrix} 0 & \omega(\tau) \\
-\Sigma_z \omega(\tau) & 0 \end{pmatrix}.
\end{align}

Applying Stokes' theorem to Eq. (\ref{Berry-phase}), we obtain the following alternative form:
\begin{align}
S_B &= \frac{N}{2} \int_0^{\beta} \!\! d \tau \, \str [\Lambda \, g^{-1} \partial_\tau g],
\end{align}
where now $g = g_{\omega(\tau)} = g(\tau, 1)$. This last form allows us, in particular, to evaluate the variation of $S_B$ upon an infinitesimal variation of the field $\Omega \to \Omega + \delta\Omega$ using integration by parts:
\begin{align}
\delta S_B &= \frac{N}{2} \int_0^{\beta} \!\! d\tau  \str \big(\Lambda  [g^{-1} \partial_\tau g, g^{-1} \delta g] \big) \nonumber \\
&= \frac{N}{4} \int_0^{\beta} \!\! d\tau \str [\Omega(\tau) \delta \Omega(\tau) \partial_\tau \Omega(\tau)].
\label{Variation-S_B}
\end{align}
This expression is very useful in the derivation of the nonlinear sigma model below.

\subsection{Modifications for lowest-weight irreps and systems of many superspins}

On the down half-links, we have superspins in the conjugate representation ${\bar R}_N$. Here, the fermions and bosons ${\bar c}_{iI}$ [see Eq. (\ref{downfb})] satisfy
\begin{align}
\llbracket {\bar c}_{iI}^{\vphantom *}, {\bar c}_{jJ}^* \rrbracket = -\delta_{ij}\delta_{IJ}.
\end{align}
The components of the conjugate superspin $\bar{\mathcal S}$ are
\begin{align}
\bar{\mathcal{S}}_{IJ} = -\sum_{i=1}^N {\bar c}_{iI} {\bar
c}_{iJ}^* - \frac{N}{2}\delta_{IJ}.
\end{align}
These generate the lowest-weight representation ${\bar R}_N$ from the vacuum $\ket{\bar 0}$, which is the lowest-weight vector with the lowest weight ${\bar \lambda}_N = - \lambda_N$. We have
\begin{align}
\bra{\bar 0} \bar{\mathcal S} \ket{\bar 0} = - \frac{N}{2} \Lambda,
\end{align}
where the matrix $\Lambda$ is the same as in Eq. (\ref{Lambda}).

The corresponding representation of the group $G$ and the coherent states are constructed as
\begin{align}
{\bar T}_g &= \exp \left[ \str(\ln g)\bar{\mathcal{S}} \right],
&\ket{\bar \omega} &= {\bar T}_{g_\omega} \ket{\bar 0}.
\end{align}
Then, we have
\begin{align}
\bra{\bar\omega} \bar{\cal S} \ket{\bar\omega} &= -\frac{N}{2} \Omega,
\end{align}
with the same matrix $\Omega$ as in Eq. (\ref{Omega}).

Finally, the action for a single conjugate superspin is
\begin{align}
S &= - S_B + \int_0^{\beta} \!\! d\tau {\cal H}[-N \Omega(\tau)/2].
\label{action-2}
\end{align}

The formulas of this appendix generalize straightforwardly to a system with many superspins. Berry phase terms for each spin add, and the rest of the action is read off from the many-superspin Hamiltonian, where one makes substitutions
\begin{align}
\mathcal{S} &\to \frac{N}{2} \Omega, & \bar{\mathcal{S}} &\to -\frac{N}{2} \Omega.
\end{align}

\section{Details of the sigma model derivation}
\label{sec:appendix-sigma model}

In this appendix, we derive the nonlinear sigma model (\ref{NLSMAction}) starting from the Hamiltonian (\ref{Hamiltonian}). To simplify equations we will suppress the integral signs as well as the supertrace sign. With this caveat in mind, using the superspin coherent states we obtain the action:
\begin{align}
S &= \sum_{l = 1,2} \big(S_B^{(l)} + S_I^{(l)}\big) + S_\perp, \\
S^{(l)}_B &= \sum_k (-1)^{k+l} S_B[\Omega_k^{(l)}],  \\
S^{(l)}_I &= -\frac{J N^2}{8} \sum_k [1 + (-1)^{k+l} \gamma_l] \big(\Omega_{k+1}^{(l)} - \Omega_k^{(l)}\big)^2, \label{S-int} \\
S_\perp &=  -\frac{J_\perp N^2}{8} \sum_k  \big(\Omega_k^{(1)} - \Omega_k^{(2)}\big)^2.
\label{S-perp}
\end{align}
Here, we used $\str \Omega^2 = \str I_{4n} = 0$.

To derive a sigma model from this action, we use the decomposition (\ref{Sierra-decomposition}). The constraints $\big(\Omega_k^{(l)}\big)^2 = I_{4n}$ lead to
\begin{align}
Q_k^2 &= I_{4n} + O(a^2),  \nonumber \\ \{Q_k, L_k\} &= \{Q_k, R_k\} = \{Q_k, M_k\} = 0,
\end{align}
and the fields $L_k$ are assumed to be small in the sense that $aL_k \ll 1$. Both $Q_k$  and $L_k$  are expected to vary little from one lattice site to the next, so they should have nice continuum limits as fields $Q(x)$ and $L(x)$.

Using the identity (\ref{Variation-S_B}), it is easy to compute the Berry phase terms:
\begin{align}
S_B[\Omega_{2k-1}^{(l)}] &= S_B[Q_k] + \frac{Na}{4} [(-1)^l (R_k - L_k) + M_k] Q_k \partial_\tau Q_k, \nonumber \\
S_B[\Omega_{2k}^{(l)}] &= S_B[Q_k] + \frac{Na}{4} [(-1)^l (R_k + L_k) - M_k] Q_k \partial_\tau Q_k.
\end{align}
Summing these with appropriate signs, we get in the continuum limit
\begin{align}
S_B = S_B^{(1)} + S_B^{(2)} = \frac{N}{2} L Q \partial_\tau Q.
\end{align}

Next we turn to the terms $S_I^{(l)}$ describing interactions along the legs of the ladder. We use the decomposition (\ref{Sierra-decomposition}) in Eq. (\ref{S-int}), replace finite differences by derivatives, and the sum over $k$ by the integral $\int \frac{dx}{2a}$ in the continuum. This gives (remember that we suppress integral signs)
\begin{align}
S_I^{(l)} &= - \frac{J N^2 a}{4} \Big[ (1 + (-1)^l \gamma_l)(\partial_x Q)^2 + 2\big((-1)^l L - M \big)^2 \nonumber \\
& \quad + 2(1 + (-1)^l \gamma_l)\big((-1)^l L - M \big) \partial_x Q  \Big].
\end{align}
Summing over $l$ we get
\begin{align}
S_I &= - \frac{J N^2 a}{2} \Big[(1 - \gamma_-)(\partial_x Q)^2 + 2L^2 + 2M^2 \nonumber \\
& \quad + 2 \gamma_+ L \partial_x Q - 2(1 - \gamma_-) M \partial_x Q\Big],
\end{align}
where $\gamma_\pm = (\gamma_1 \pm \gamma_2)/2$.

The remaining term in the action, $S_\perp$, describing the interaction of superspins along the rungs of the ladder is treated similarly and gives in the continuum
\begin{align}
S_\perp &= - \frac{J_\perp N^2 a}{2} (R^2 + L^2).
\end{align}
There are two noticeable features of this action. First of all, the field $R$ is completely decoupled from all other fields. Secondly, its mass is proportional to $J_\perp$, so we may expect the decomposition (\ref{Sierra-decomposition}) to become less and less meaningful and useful as we approach the point of decoupled chains $J_\perp = 0$. The masses of other fields to be integrated out ($M$ and $L$) are set by $J$, and so they remain finite even at $J_\perp = 0$.

For now we simply disregard the $R$ field and integrate out the $M$ field, which leads to
\begin{align}
S &= \frac{N}{2} L Q \partial_\tau Q - \frac{J N^2 a}{4}(1 - \gamma_-^2)(\partial_x Q)^2 \nonumber \\
&\quad - J N^2 a \big(\mu L^2 + \gamma_+ L \partial_x Q \big),
\end{align}
where we have introduced the parameter
\begin{align}
\mu \equiv 1 + \frac{J_\perp}{2J}.
\end{align}
Then we integrate out the $L$ field:
\begin{align}
S &= - \frac{J N^2 a}{4\mu}(\mu - \mu \gamma_-^2 - \gamma_+^2) (\partial_x Q)^2 \nonumber \\
& \quad - \frac{1}{16 J a \mu} (\partial_\tau Q)^2 - \frac{N \gamma_+}{4\mu} Q \partial_\tau Q \partial_x Q,
\end{align}
Now, we only need to rescale the time variable $\tau \to \tau/\lambda$ to make the coefficients of the quadratic terms equal. This leads to the isotropic sigma model (\ref{NLSMAction}), with coefficients given by Eq. (\ref{conductivities-two-leg-ladder}).


\begin{references}
\bibitem{footnote} This is in full analogy with previous studies of single-channel network models [\onlinecite{DHLee,GruzReadSach}].
\end{references}
\end{document}